\documentclass[fleqn,usenatbib]{mnras}

\usepackage{newtxtext,newtxmath}

\usepackage[T1]{fontenc}

\usepackage{ragged2e}

\usepackage{lineno}
\usepackage{booktabs}
\usepackage{graphicx}

\usepackage{xspace}
\usepackage{float}
\usepackage{hyperref}
\usepackage{color, colortbl}
\usepackage{siunitx}

\usepackage{scalerel}
\usepackage{tikz} 
\usetikzlibrary{svg.path}

\definecolor{orcidlogocol}{HTML}{A6CE39}
\tikzset{
  orcidlogo/.pic={
    \fill[orcidlogocol] svg{M256,128c0,70.7-57.3,128-128,128C57.3,256,0,198.7,0,128C0,57.3,57.3,0,128,0C198.7,0,256,57.3,256,128z};
    \fill[white] svg{M86.3,186.2H70.9V79.1h15.4v48.4V186.2z}
                 svg{M108.9,79.1h41.6c39.6,0,57,28.3,57,53.6c0,27.5-21.5,53.6-56.8,53.6h-41.8V79.1z M124.3,172.4h24.5c34.9,0,42.9-26.5,42.9-39.7c0-21.5-13.7-39.7-43.7-39.7h-23.7V172.4z}
                 svg{M88.7,56.8c0,5.5-4.5,10.1-10.1,10.1c-5.6,0-10.1-4.6-10.1-10.1c0-5.6,4.5-10.1,10.1-10.1C84.2,46.7,88.7,51.3,88.7,56.8z};
  }
}

\newcommand\orcid[1]{\href{https://orcid.org/#1}{\mbox{\scalerel*{
\begin{tikzpicture}[yscale=-1,transform shape]
\pic{orcidlogo};
\end{tikzpicture}
}{|}}}} 

\usepackage{graphicx}	
\usepackage{amsmath}	

\newcommand{\tess}{{\it TESS}}
\newcommand{\thisstar}{TIC 434398831} 
\newcommand{\thisstarb}{TIC 434398831\,b}
\newcommand{\thisstarc}{TIC 434398831\,c}
\newcommand{\radb}{$R_b = 3.51_{-0.21}^{+0.22}\,\mathrm{R_\oplus}$}
\newcommand{\shortradb}{$R_b = 3.51\,\mathrm{R_\oplus}$}
\newcommand{\radc}{$R_c = 5.63_{-0.28}^{+0.29}\,\mathrm{R_\oplus}$}
\newcommand{\shortradc}{$R_c = 5.63\,\mathrm{R_\oplus}$}
\newcommand{\perb}{$P_b = 3.685504_{-0.000011}^{+0.000012}$ days}
\newcommand{\shortperb}{$P_b = 3.69$ days}
\newcommand{\perc}{$P_c = 6.210291_{-0.000020}^{+0.000013}$ days}
\newcommand{\shortperc}{$P_c = 6.21$ days}

\newcommand{\shortmstar}{$M_\star = 0.99\, \mathrm{M_\odot}$}

\newcommand{\shortrstar}{$R_\star = 0.91\, \mathrm{R_\odot}$}
\newcommand\mysim{\mathord{\sim}}
\newcommand{\age}{$61\pm6$\, Myr}

\newcommand{\MITKavli}{Department of Physics and Kavli Institute for Astrophysics and Space Research, Massachusetts Institute of Technology, Cambridge, MA 02139, USA}

\title[\thisstar]{A transiting multi-planet system in the 61 million year old association Theia 116} 

\author[Vach, et al.]{\parbox{\textwidth}
{Sydney Vach\orcid{0000-0001-9158-9276},$^{1}$\thanks{E-mail: \texttt{sydney.vach@unisq.edu.au}}
George Zhou\orcid{0000-0002-4891-3517},$^{1}$
Chelsea X. Huang\orcid{0000-0003-0918-7484},$^{1}$
Andrew W. Mann\orcid{0000-0003-3654-1602},$^2$
Madyson G. Barber\orcid{0000-0002-8399-472X},$^{2}$\thanks{NSF Graduate Research Fellow}
Allyson Bieryla,$^{1,3}$
David W. Latham\orcid{0000-0001-9911-7388},$^{3}$
Karen A. Collins\orcid{0000-0001-6588-9574},$^{3}$
James G. Rogers\orcid{0000-0001-7615-6798},$^4$
Luke G. Bouma\orcid{0000-0002-0514-5538},$^{5}$\thanks{51 Pegasi b Fellow}
Stephanie T. Douglas\orcid{0000-0001-7371-2832},$^6$
Samuel N. Quinn\orcid{0000-0002-8964-8377},$^{3}$
Tyler R. Fairnington\orcid{0000-0002-0692-7822},$^{1}$
Joachim Kr\"uger\orcid{0009-0003-3841-5383},$^{1}$
Avi Shporer\orcid{0000-0002-1836-3120},$^7$
Kevin I. Collins\orcid{0000-0003-2781-3207},$^{8}$
Gregor Srdoc,$^{9}$
Richard P. Schwarz\orcid{0000-0001-8227-1020},$^{3}$
Howard M. Relles\orcid{0009-0009-5132-9520},$^{3}$
Khalid Barkaoui\orcid{0000-0003-1464-9276},$^{10,11,12}$
Kim K. McLeod\orcid{0000-0001-9504-1486},$^{13}$
Alayna Schneider\orcid{0009-0005-4771-4654},$^{13}$
Norio Narita\orcid{0000-0001-8511-2981},$^{12,14,15}$
Akihiko Fukui\orcid{0000-0002-4909-5763},$^{12,14}$
Steve~B.~Howell\orcid{0000-0002-2532-2853},$^{16}$
Colin Littlefield\orcid{0000-0001-7746-5795},$^{17}$
Sarah Deveny\orcid{0009-0002-9833-0667},$^{17,16}$
Ramotholo Sefako\orcid{0003-3904-6754},$^{18}$
William Fong,$^7$
Ismael Mireles\orcid{0000-0002-4510-2268},$^{19}$
Guillermo Torres,$^3$
George R. Ricker\orcid{0000-0003-2058-6662},$^{7}$
Sara Seager\orcid{0000-0002-6892-6948},$^{7,11,20}$
Joshua N. Winn\orcid{0000-0002-4265-047X}$^{21}$
}
}

\date{Accepted XXX. Received YYY; in original form ZZZ}

\pubyear{2024}

\begin{document}
\label{firstpage}
\pagerange{\pageref{firstpage}--\pageref{lastpage}}
\maketitle

\begin{abstract}
Observing and characterizing young planetary systems can aid in unveiling the evolutionary mechanisms that sculpt the mature exoplanet population. As an all-sky survey, NASA's Transiting Exoplanet Survey Satellite (\tess) has expanded the known young planet population as it has observed young comoving stellar populations. This work presents the discovery of a multiplanet system orbiting the 61$\pm6$ Myr old G4V star \thisstar\ (\shortmstar, \shortrstar, $T_\mathrm{eff} = 5638\,$K, $T\mathrm{mag} = 11.31$) located in the Theia 116 comoving population. We estimate the population's age based on rotation periods measured from the \tess\ light curves, isochrone fitting, and measurements of lithium equivalent widths in the spectra of Theia 116 members. The \tess\ FFI light curves reveal a mini-Neptune (\shortradb, \shortperb) and super-Neptune (\shortradc, \shortperc) with an orbital period ratio slightly larger than 5:3. Follow-up observations from CHEOPS and ground-based telescopes confirm the transits of \thisstarb\ and c, and constrain their transit times. We explore the potential mass-loss histories of the two planets in order to probe possible initial conditions of the planets immediately after formation.

\end{abstract}




\section{Introduction}

Young planetary systems can act as make-shift time machines, allowing us to observe universal evolutionary processes that carved out the mature exoplanet population. 
The first few 100 million years after formation are thought to be highly important in the evolution of planetary systems. 
Gravitational interactions sculpt the orbital properties of these newborn planetary systems \citep[e.g.,][]{Ida:2004}. Interactions between the young star and the planet's atmosphere result in atmospheric erosion leading to a change in the mass and radius of the young planet \citep{Lopez:2013, owen:2013, Owen:2017}. As a result, the demographics of the youngest exoplanet population vary from its mature counterpart \citep{Fernandes:2023, Vach:2024}, as they are still undergoing these processes.

While detecting and characterizing young planetary systems enables us to probe these important universal processes, young planetary systems can be difficult to detect, in part due to the challenges in determining precise ages for individual stars. Luckily, when stars still reside in the comoving populations in which they were born, we are able to estimate the age of the entire population of stars with a higher precision. While NASA's Kepler mission was able to detect a handful of young planetary systems \citep[e.g., ][]{Bouma:2022, Barber:2022}, the \textit{K2} mission \citep{k2} targeted well-characterized clusters, unveiling a collection of young planetary systems \citep[e.g., ][]{David:2016,Mann2016,David:2019} spread across the solar neighborhood. These provided the first glimpses of the apparent differences between the young and mature planet populations. 

In the age of \emph{Gaia} \citep[][]{GaiaDR3}{}{}, kinematics and clustering analyses of \emph{Gaia} spatial velocities and positions have unveiled previously unknown young, comoving populations \citep[][]{Kounkel:2019, Moranta:2022}{}{}, identifying new targets to search for young transiting exoplanets across the entire sky. As an all-sky survey, NASA's Transiting Exoplanet Survey Satellite \citep[\tess;][]{TESS}{}{} has now observed over 95\% of the sky, encompassing both well-characterized stellar clusters and newly identified comoving populations, and allowing for a near-doubling of the number of known young transiting exoplanets \citep[e.g.][]{Newton2019,Rizzuto2020,Plavchan2020,Tofflemire2021}{}{}. 

In this work, we present the discovery and characterization of the young multi-planet system \thisstar\, located in the recently identified comoving population Theia 116 \citep[][]{Kounkel:2019}{}{}. Originally discovered by \cite{Vach:2024}, \thisstar\ hosts the mini-Neptune \thisstarb\ (\shortradb, \shortperb), and the super-Neptune \thisstarc\ (\shortradc, \shortperc). In section~\ref{sec:obs}, we present our \tess, \textit{CHEOPS}, and ground-based observations of \thisstar. Section~\ref{sec:star} describes our stellar and cluster characterization, including our cluster age calculation using rotation periods, lithium absorption strengths, and fitting isochrones to the spectral energy distributions (SED) of the members of Theia 116. We describe our global model in Section~\ref{sec:global_model} and discuss our investigations into possible false-positive scenarios in Section~\ref{sec:fp}. We present our discussion of the mass loss history of the system in Section~\ref{sec:discussion}.

\section{Observations}\label{sec:obs}

\subsection{\emph{TESS} Photometry}
NASA's all-sky photometric survey spacecraft, \tess, was launched in April 2018 with the primary objective of discovering transiting planets around nearby, bright stars. \tess\ uses 4 cameras, each composed of 4 CCDs, to monitor a $96^\circ\times24^\circ$ strip of the sky, called a sector, for a stare duration of $\mysim27$ days. Preselected target stars have their light curves sampled at 2-minute averages of 2-second exposures, while the entire field of view is sampled in the Full Frame Images (FFIs). FFIs were taken at 30-minute, 10-minute, and 200-second cadences in the primary, first extended, and second extended missions respectively. 

\thisstar\ was observed in the \tess\ FFIs across Sectors 6, 33, 43, 44, 45, 71, and 72. Data from Sectors 6, 33, 43, 44, 45, 71, and 72 were processed by the MIT Quick Look Pipeline \citep[QLP; ][]{qlp2020a, qlp2020b}{}{}, and made available via the Mikulski Archive for Space Telescopes (MAST). \thisstar\ was identified as a planet candidate host in an independent survey for transiting planets around young stars in comoving populations observed by \tess\, \citep[][]{Vach:2024}{}{}. In short, a magnitude-limited search of kinematically associated, comoving populations with isochrone ages $<200$ Myr with \tess\ for transiting planet signals was performed. After detrending for stellar activity following \citet{Vanderburg:2019}, a Box Least Squares \citep[BLS; ][]{kovacs:2002}{}{} search was used to identify transit signals. Identified signals were required to have a signal-to-pink noise $\geq 8$ to be considered a threshold crossing event (TCE). TCEs were then subjected to a vetting procedure, including per-pixel analysis, centroid analysis, odd-even analysis, and visual vetting to ensure the transit event was on target and consistent with a transiting planet event. 

\begin{figure*}
    \centering
    \includegraphics[width=0.71\linewidth]{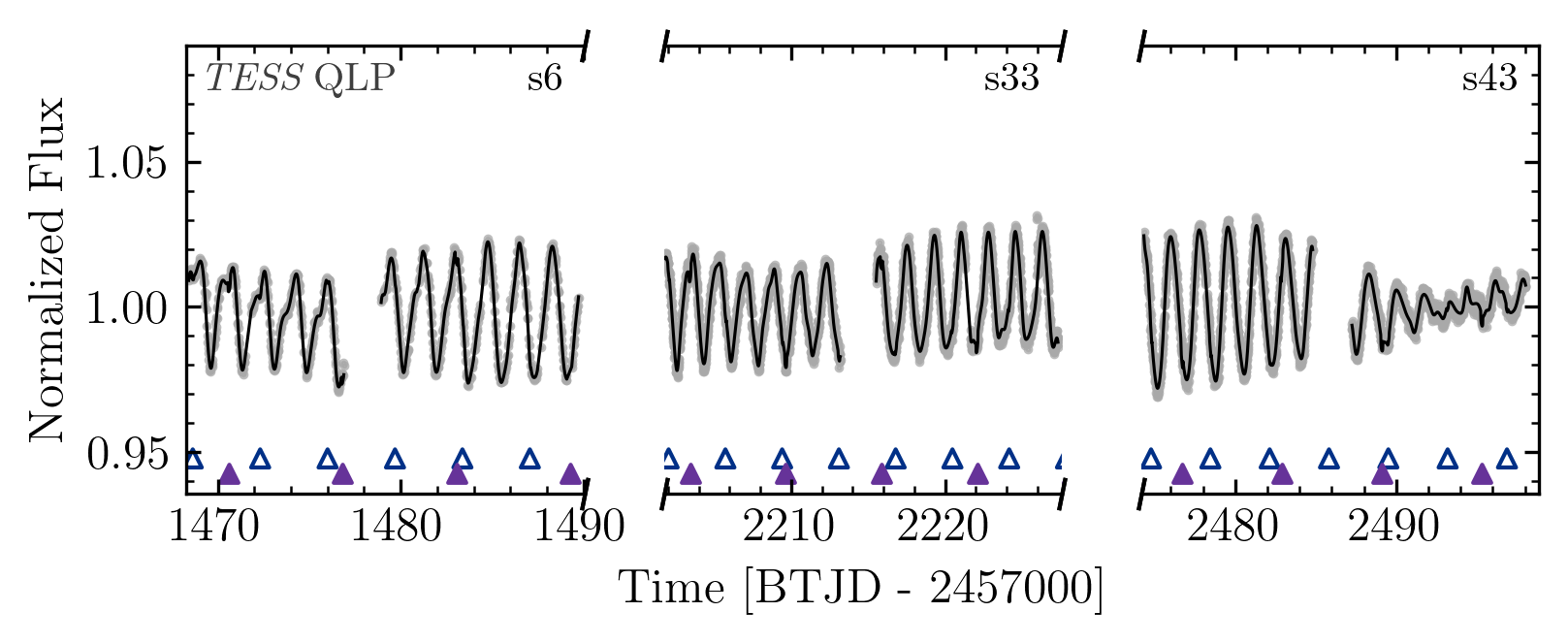}\\
    \includegraphics[width=0.9\linewidth]{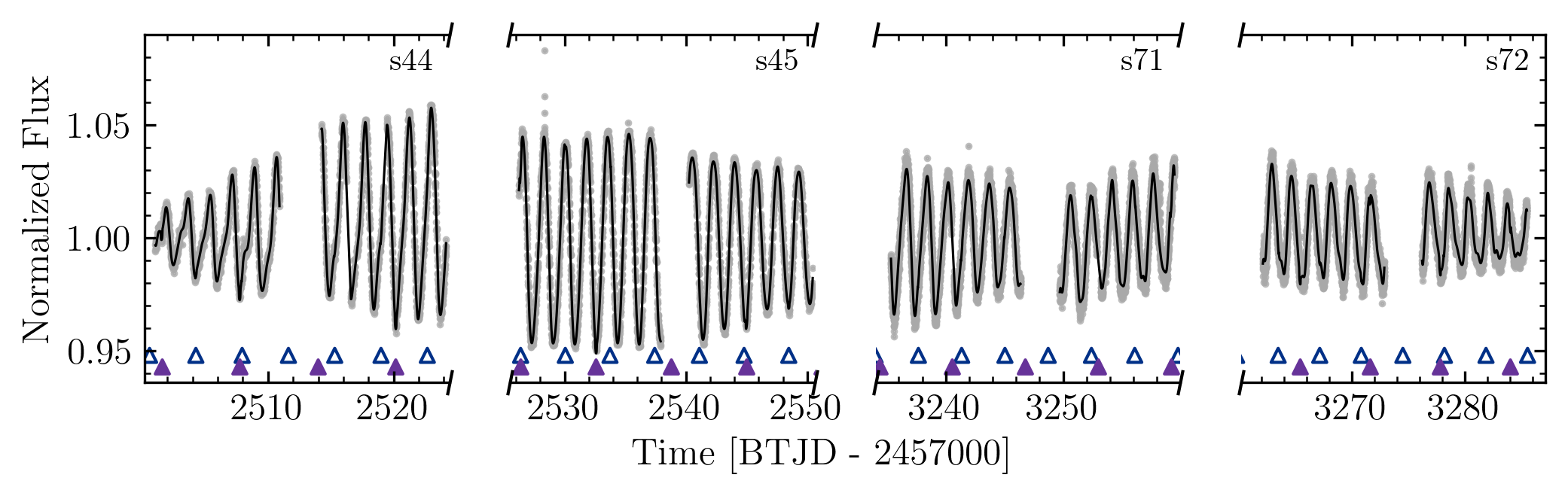}\\
    \includegraphics[width=0.7\linewidth]{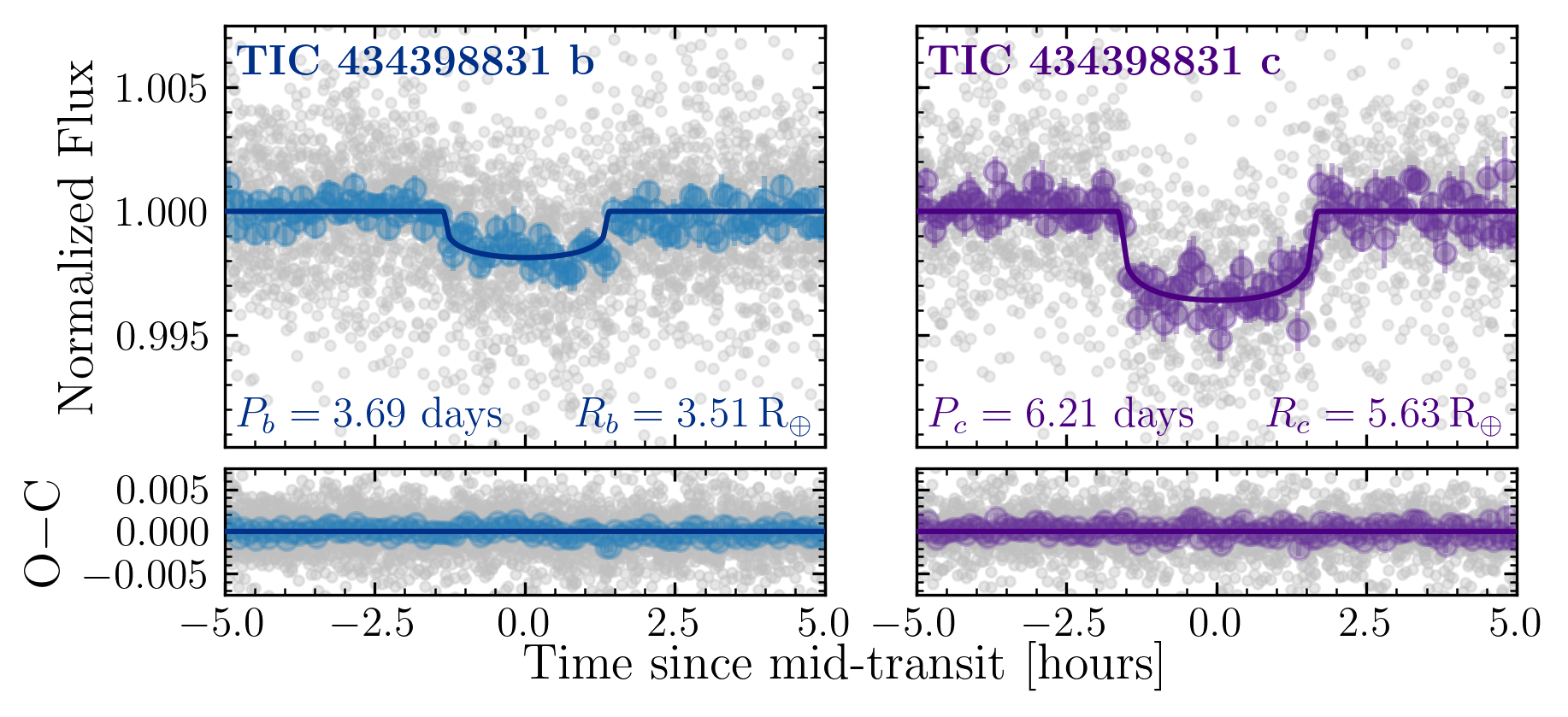}
    \caption{\tess\ QLP light curves of \thisstar\ (top and middle panels). \tess\ observed \thisstar\ across Sectors 6, 33, 43, 44, 45, 71, and 72 in the FFIs. We modeled the photometric variability induced by stellar activity combined with the best-fit transit model, shown in black. Transits of \thisstarb\ (blue empty triangles) and \thisstarc\ (purple triangles) are marked beneath the full \tess\ light curves. The phase folded transits and residuals of \thisstarb\ (left) and c (right) are plotted in the bottom panel. Overlaid is the best-fit transit model for \thisstarb\ and c.}
    \label{fig:tess_lc}
\end{figure*}

In the FFI light curves for \thisstar, a 6.21 day periodic event, \thisstarc, triggered a TCE with a signal-to-pink noise of 15.7. This signal passed all vetting procedures, and was advanced to planet-candidate status. \thisstar\ was then checked for multiplicity by masking the 6.21 day periodic event and rerunning the planet search. This identified an additional 3.68 day TCE, with a signal-to-pink noise 11.8, \thisstarb. We present the full \tess\ light curve and the phase-folded \tess\ transits of \thisstarb\, and c in Figure~\ref{fig:tess_lc}. Our global model best-fit parameters are presented in Section~\ref{sec:global_model}. 

\subsection{\emph{CHEOPS} Photometry}\label{sec:cheops}

\begin{figure*}
    \centering
    \includegraphics[width=\linewidth]{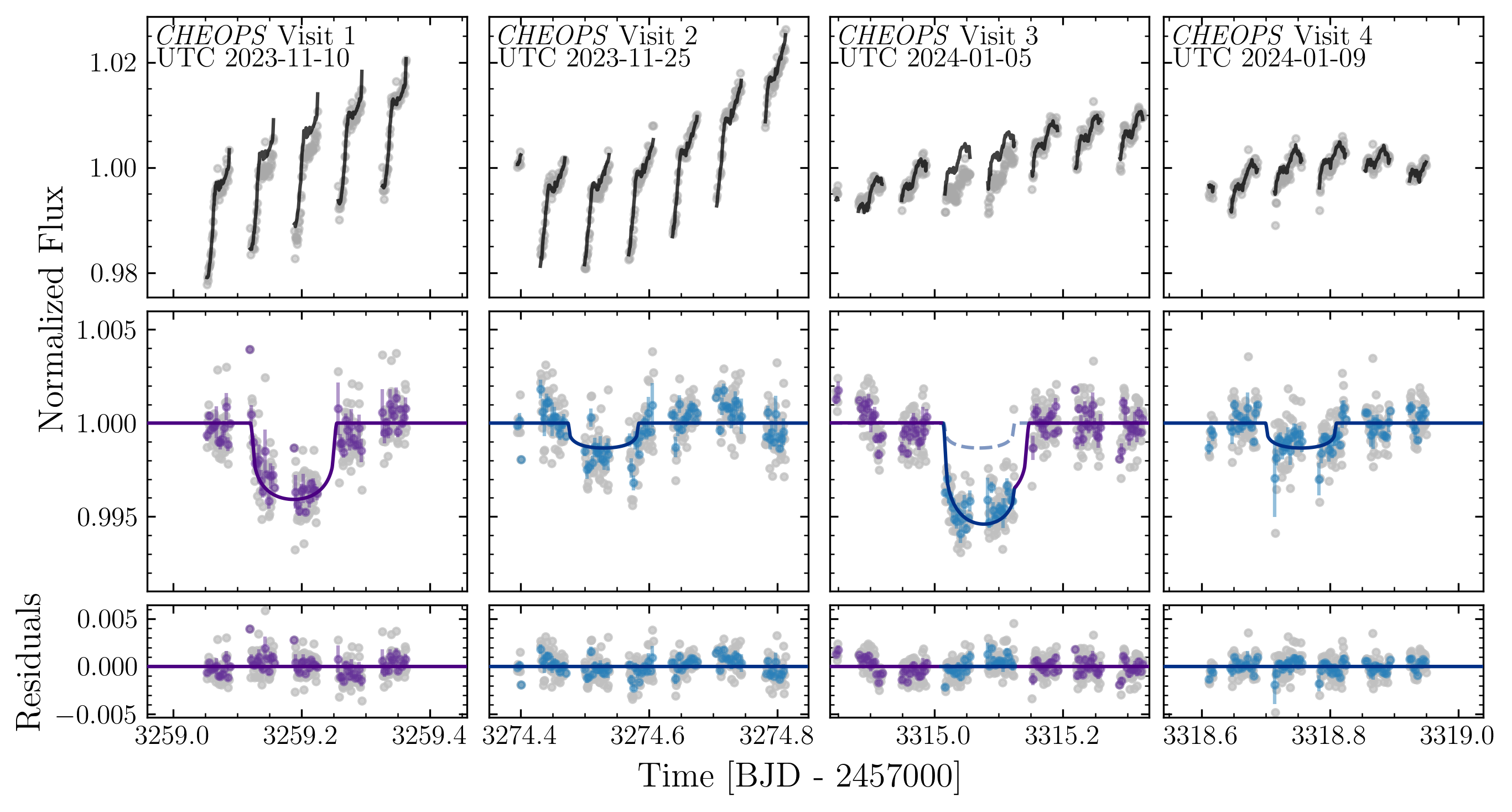}
    \caption{\textit{CHEOPS} light curves from 4 visits of \thisstar. The top panel shows the raw \textit{CHEOPS} data (grey) and our best-fit trend model (black). We plot the detrended light curves with our best-fit transit models overlaid in the middle panel. Visits 2 and 3 observed full transits of \thisstarb\ (blue). A full transit of \thisstarc\ (purple) was observed in visit 1. A double transit event of \thisstarb\ and c was observed in visit 3. We plot the observations during the transit of \thisstarb\ in blue, in addition to the transit of \thisstarc\ shown in purple.
    Residuals are plotted in the bottom panel. }
    \label{fig:cheops}
\end{figure*}

We obtained follow-up observations with ESA's CHaracterizing ExOPlanets Satellite \citep[\textit{CHEOPS};][]{Cheops}{}{} of \thisstarb\ and c as part of the \textit{CHEOPS} Guest Observing Program (AO4-002, PI: Vach). \textit{CHEOPS} observes a $17'\times17'$ field of view with a 32 cm telescope, orbiting on the day-night terminator every $\mysim90$ min. 

\begin{figure}
\includegraphics[width=0.9\linewidth]{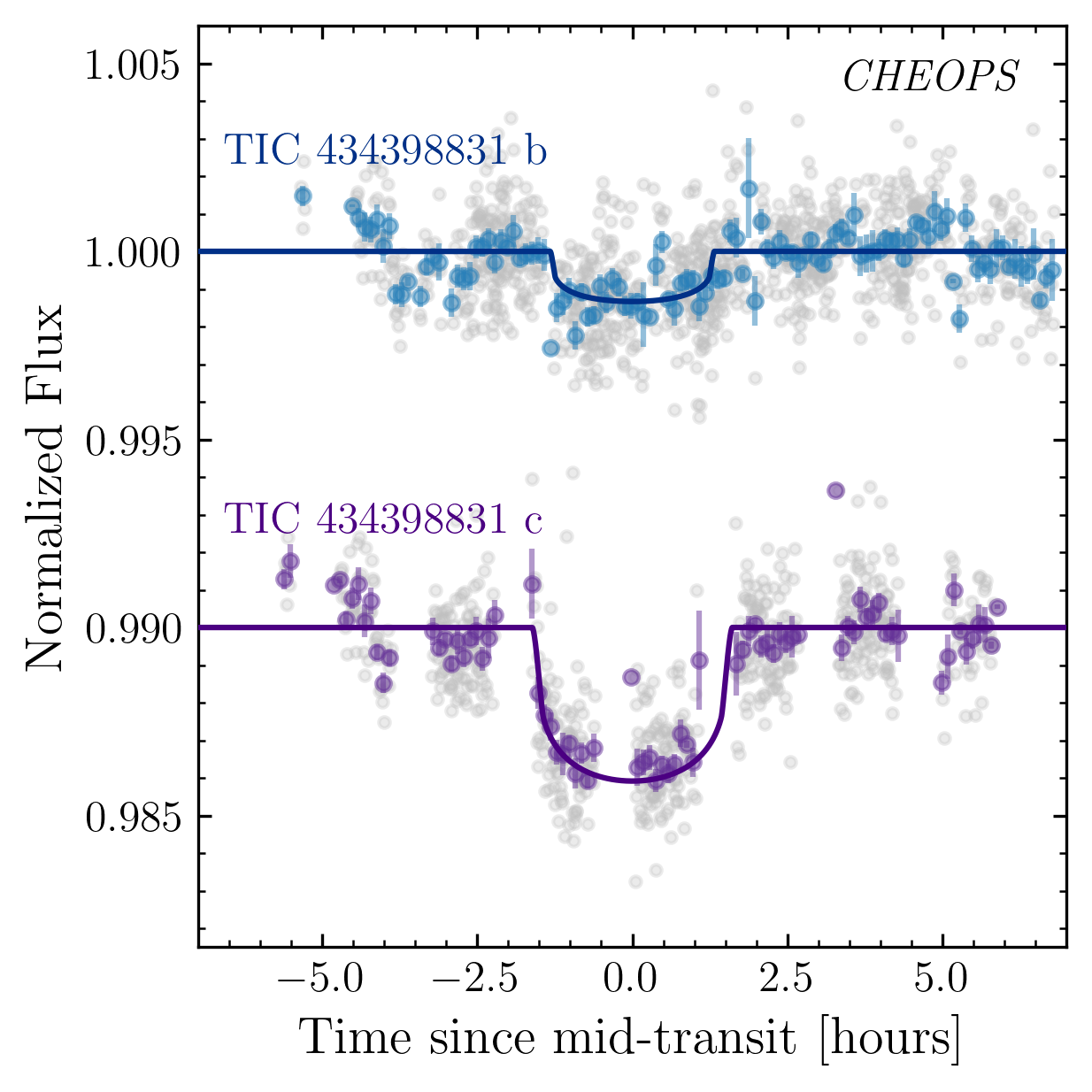}
    \caption{Phase folded \textit{CHEOPS} light curves (grey) of \thisstarb\ (blue) and c (purple). We overlay our best-fit transit models. }
    \label{fig:phased_cheops}
\end{figure}

\thisstar\ was observed across 4 visits, spanning 26 orbits at 60 s exposure times. Visit 1 (UTC 2023-11-10) observed a full transit of \thisstarb\, across 5 orbits, for a total duration of 447 minutes. Visits 2 (UTC 2023-11-25) and 4 (UTC 2024-01-09) observed full transits of \thisstarc, spanning 7 orbits (600 minutes) and 6 orbits (486 minutes) respectively. Visit 3 (UTC 2024-01-05) observed a double transit event of both \thisstarb\ and c, across 8 orbits, with a duration of 687 minutes. 

Observations from the 4 visits were reduced via the \textit{CHEOPS} pipeline \citep[][]{Hoyer:2020}{}{}. For subsequent analyses, we made use of the light curves extracted from the \textit{CHEOPS} default aperture for each visit. We masked out any data with bad quality flags and removed $>5\sigma$ outliers with an iterative sigma clipping. Spacecraft systematics take place on the time scale of the \textit{CHEOPS} orbit. Following the trend model described in \textsc{Pycheops} \citep{pycheops}, we modeled for systematics using background flux, aperture contamination, target smearing, roll angle, and centroid $X$ and $Y$ coordinates. We simultaneously detrended the raw \textit{CHEOPS} data in our global model in order to propagate instrument uncertainties into our best-fit parameters by fitting a least-squares model to the light curve residuals. The raw and detrended \textit{CHEOPS} light curves for \thisstarb\ and c are presented in Figure~\ref{fig:cheops}. The combined phased folded \textit{CHEOPS} light curves are shown in Figure~\ref{fig:phased_cheops}

\subsection{Ground-based Photometry}
\begin{figure}
    \includegraphics[width=0.9\linewidth]{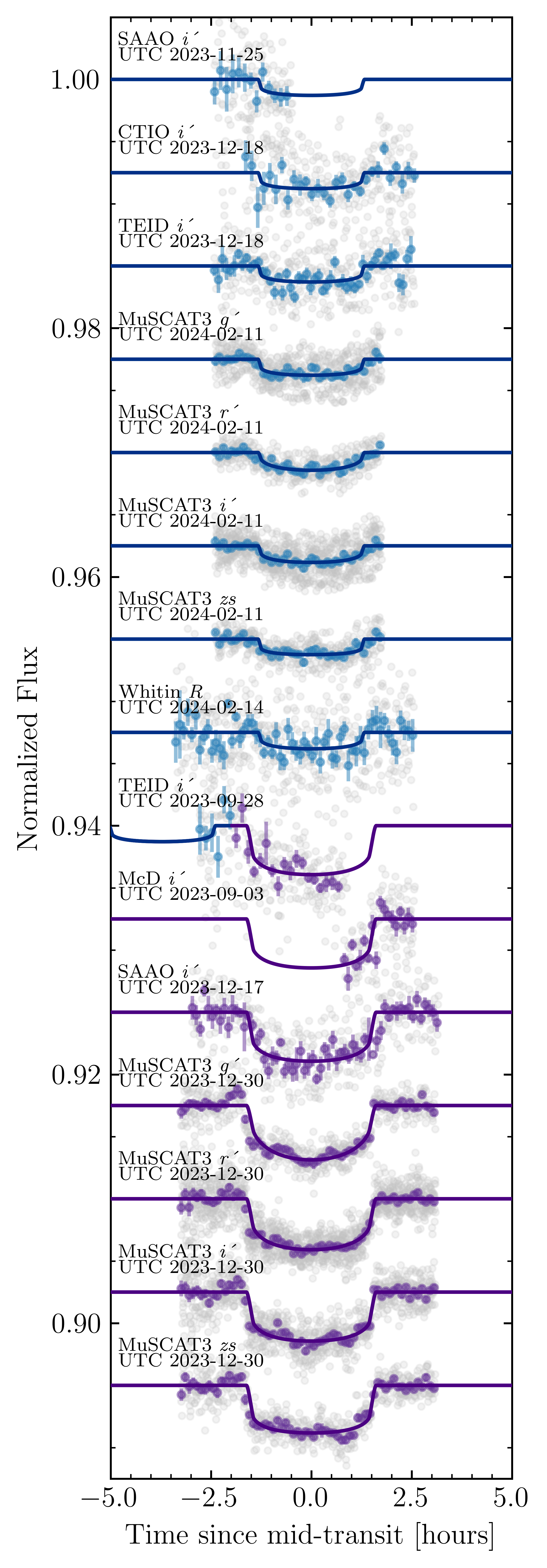}
    \caption{Ground-based photometric follow-up of \thisstarb\ (blue) and c (purple). Our best-fit transit models are overlaid.}
    \label{fig:lco_lc}
\end{figure}

We obtained ground-based photometry as part of the \tess\ Follow-up Observing Program \citep[TFOP;][]{collins:2019}\footnote{https://tess.mit.edu/followup}. We used the {\tt TESS Transit Finder}, which is a customized version of the {\tt Tapir} software package \citep{Jensen:2013}, to schedule our transit observations. A summary of our ground-based photometric observations is presented in Table~\ref{tab:lco}, with the light curves shown in Figure~\ref{fig:lco_lc}. The ground-based observations confirm the transits of \thisstarb\ and c are on target to within 3.1'', and clear the field for nearby eclipsing binary (NEB) scenarios $<2.5$'. We made use of the raw ground-based photometry and simultaneously detrended the transit observations in our global model. Theses observations are made publically available through Exoplanet Follow-up Observing Program (ExoFOP).\footnote{\href{https://exofop.ipac.caltech.eduTESStarget.php?id=434398831 
}{https://exofop.ipac.caltech.eduTESStarget.php?id=434398831 
}}

\subsubsection{LCOGT Sinistro}

We observed \thisstar\ with 8 observations by the Las Cumbres Observatory Global Telescope Network \citep[LCOGT; ][]{Brown:2013}{}{}{} of 1-meter telescopes. LCOGTs' $1024\times1024$ SINISTRO cameras observe a 13’ x 13’ field of view, with a 0.778” pixel scale. 

LCO-McDonald Observatory (LCO-McD), Texas, USA, captured an egress of \thisstarc\, on UTC 2023-09-03. LCO-Teide Observatory (LCO-TEID), Tenerife, Spain, observed an ingress of \thisstarc\, and an egress of \thisstarb\, on UTC 2023-09-28. An additional full transit of \thisstarb\, was observed on UTC 2023-11-25. LCO-Teide and LCO-Cerro Tololo Inter-American Observatory (LCO-CTIO), Cerro Tololo, Chile, simultaneously observed a full transit of \thisstarb\, on UTC 20233-12-18. LCO-South African Astronomical Observatory (LCO-SAAO) observed an ingress of \thisstarb\, on UTC 2023-11-25, and a full transit of \thisstarc\, on UTC 2023-12-17. All observations were conducted in the \textit{i'} band. The images were calibrated with the LCOGT \texttt{BANZAI} pipeline \citep[][]{McCully:2018}{}{}. Light curves were then extracted with \texttt{AstroImageJ} \citep[][]{Collins:2017}{}{}. 

We used all LCOGT Sinistro observations in our global modeling (see Section~\ref{sec:global_model}), with the exception of the LCO-TEID observations of \thisstarb, UTC 2023-11-25 and UTC 2024-03-11, which experienced large scatter due to poor weather and systematics respectively.

\subsubsection{LCOGT MuSCAT3}

The LCOGT Multicolor Simultaneous Camera for studying Atmospheres of Transiting exoplanets \citep[MuSCAT3;][]{Narita:2020} imager located on the 2-m Faulkes Telescope North at Haleakal\={a} Observatory, Haleakal\={a}, Hawai'i observed a full transit of both \thisstarb\, and c. MuSCAT3 has a pixel scale of 0.27" covering a 9.1'$\times$9.1' field of view.

MuSCAT3 observed a full transit of \thisstarb\, (UTC 2024-02-11) and a full transit of \thisstarc\, (UTC 2023-12-30) simultaneously in the Sloan $g'$, $r'$, $i'$, and  $z_s$ bands. The images were calibrated using the standard LCOGT \texttt{BANZAI} pipeline \citep[][]{McCully:2018}{}{}, and the light curves were extracted using \texttt{AstroImageJ} \citep[][]{Collins:2017}{}{}. We made use of all MuSCAT3 observations in our global modeling procedure. 

\subsubsection{Whitin Observatory}

We observed a full transit of \thisstarb\, (UTC 2024-02-14) in the $R$ band with the 0.7-m PlaneWave CDK700 telescope located at the Wellesley College Whitin Observatory, Massachusetts, USA. The images were obtained with an SBIG camera with a pixel scale of 1.08" covering an 18.4'$\times$18.4' field of view. We reduced the data and extracted the light curves with \texttt{AstroImageJ} \citep[][]{Collins:2017}{}{}. The Whitin transit observation of \thisstarb\ was used in our global model.


\begin{table}
	\caption{Ground-based photometry of \thisstar.}
	\label{tab:lco}
	\begin{tabular}{lcccc} 
		\hline
		\,\,\,Instrument & Date (UTC) & Duration & Aperture &Filter\\
		\hline
		\textbf{TIC 434398831 b}\\
            \,\,\,LCO-TEID 1.0\,m& 2023-09-28 & egress & 3.1'' & \textit{i'} \\
		\,\,\,LCO-SAAO 1.0\,m&2023-11-25  & ingress & 5.1'' &\textit{i'}\\
		\,\,\,LCO-TEID 1.0\,m$^*$& 2023-11-25  & full & 5.1''& \textit{i'}\\
            \,\,\,LCO-CTIO 1.0\,m&2023-12-18  & full & 4.3''& \textit{i'}\\
            \,\,\,LCO-TEID 1.0\,m&2023-12-18  & full & 4.3'' &\textit{i'}\\
            \,\,\,MuSCAT3 2.0\,m& 2024-02-11  & full & 5.3''&\textit{i'}\\
            \,\,\,MuSCAT3 2.0\,m& 2024-02-11  & full & 5.6''&\textit{g'}\\
            \,\,\,MuSCAT3 2.0\,m& 2024-02-11  & full & 5.3''&\textit{r'}\\
            \,\,\,MuSCAT3 2.0\,m& 2024-02-11  & full & 5.3''&$z_s$\\
            \,\,\,Whitin 0.7\,m& 2024-02-14  & full & 5.5'' &\textit{R}\\
            \,\,\,LCO-TEID 1.0\,m$^*$& 2024-03-11  & full & 5.5'' &\textit{i'}\\
            \textbf{TIC 434398831 c}\\
            \,\,\,LCO-McD 1.0\,m& 2023-09-03 & egress & 4.3'' & \textit{i'} \\
            \,\,\,LCO-TEID 1.0\,m& 2023-09-28 & ingress & 3.1''& \textit{i'} \\
            \,\,\,LCO-SAAO 1.0\,m& 2023-12-17 & full & 5.8''&\textit{i'} \\
            \,\,\,MuSCAT3 2.0\,m& 2023-12-30 & full & 4.7'' &\textit{i'} \\
            \,\,\,MuSCAT3 2.0\,m& 2023-12-30 & full & 4.7'' &\textit{g'} \\
            \,\,\,MuSCAT3 2.0\,m& 2023-12-30 & full & 4.7'' &\textit{r'} \\
            \,\,\,MuSCAT3 2.0\,m& 2023-12-30 & full & 4.7'' &$z_s$ \\
		\hline
	\end{tabular}
 $^*$\textit{Not included in global model due to systematics.}
\end{table}

\subsection{High-Resolution Imaging}\label{sec:hri}
\begin{figure}
    \centering
    \includegraphics[width=\linewidth]{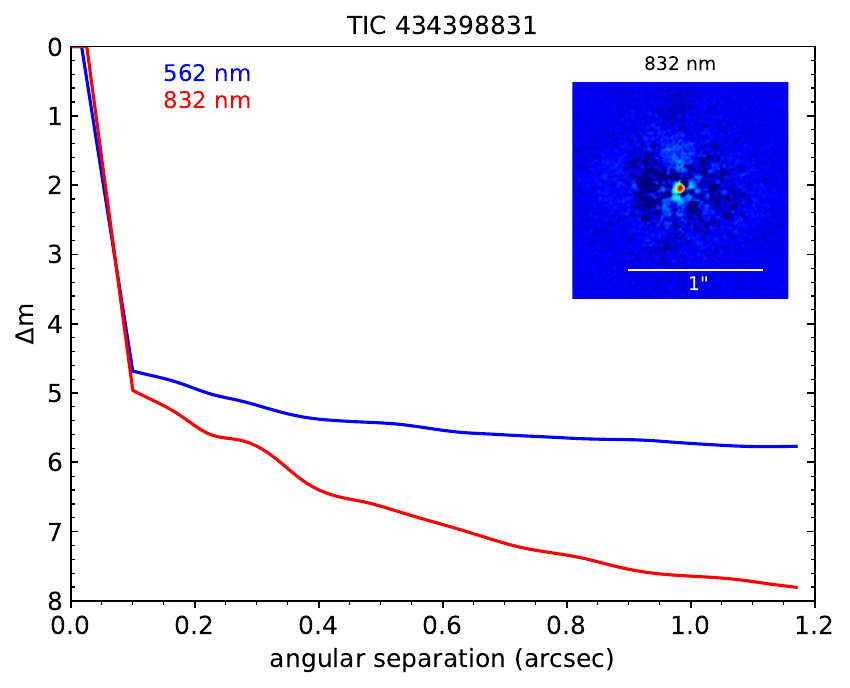}
    \caption{High-resolution image of \thisstar\ with 'Alopeke on Gemini North. No secondary source was detected. Our $5\sigma$ contrast curves for the red and blue arm are plotted.}
    \label{fig:hri}
\end{figure}
To search for visual companions we obtained high-resolution images of \thisstar. We observed \thisstar\ using the red (562 nm) and blue (832 nm) arm of the `Alopeke speckle imager. `Alopeke is located on the 8-meter Gemini North Telescope \citep{Scott:2021}, with a field of view of $2.''5\times2.''5$ and a pixel scale of 0.''01 per pixel.

We observed \thisstar\ with `Alopeke on 2024 September 27. Images were reduced using a custom pipeline \citep{Horch,Howell:2011} developed by the `Alopeke group. No secondary sources were detected around \thisstar. Our observations achieved a contrast ratio of $\Delta$6.63 mag at 0.5'' in the red and $\Delta$5.43 mag at 0.5'' in the blue. Our resulting $5\sigma$ contrast curves are presented in Figure~\ref{fig:hri}.

\subsection{Reconnasance Spectroscopy}\label{sec:spec}

We obtained a set of spectroscopic observations to characterize the spectroscopic properties of the host star and to search for signatures of astrophysical false positive scenarios. We obtained seven epochs of observations with the Tillinghast Reflector Echelle Spectrograph (TRES; \citealt{gaborthesis}) on the 1.5\,m reflector at the Fred Lawrence Whipple Observatory, Arizona. TRES is a high-resolution echelle spectrograph with a resolving power of $R \approx 44,000$, with a wavelength coverage of $3850-9100\,$\AA. Observing strategy and spectral extraction techniques are outlined in \citet{2012Natur.486..375B}. For each spectral epoch, a sequence of three consecutive exposures are obtained and combined with sigma clipping applied to mitigate the effects of cosmic ray artefacts. Wavelength calibration is provided via bracketing thorium-argon lamp exposures before and after the target exposure. Spectroscopic atmospheric parameters are derived from each spectrum via the Stellar Parameter Classification (SPC) pipeline as per \citet{2012Natur.486..375B}, via a cross correlation between the observed spectrum and a grid of spectral templates from the \citet{1992IAUS..149..225K} model atmospheres. An interpolation over the cross correlation function peak heights, as a function of the stellar atmosphere parameters, provides the adopted spectral parameters. Uncertainties in the atmosphere parameters from SPC are estimated based on the scatter in the parameters for a test population that was also classified via the Spectroscopy Made Easy \citep{1996A&AS..118..595V}. Relative velocities were derived from a cross-correlation between each epoch spectrum and a template constructed from all TRES observations of the target as per \citet{2012ApJ...745...80Q}. Cross-correlation functions are derived over $\sim 25$ of the best orders, least affected by telluric and major Balmer lines. The cross-correlation functions from these orders are added and fit via a Gaussian to derive the relative RV from each epoch. The radial velocity uncertainties are determined from via $\sigma_\mathrm{rv} = \mathrm{RMS} / \sqrt{N}$, where RMS is the root mean square scatter of the relative velocities from each order, and $N$ is the number of orders used in the analysis. These are presented in Table~\ref{tab:tresrv}.

We find \thisstar\ to have an effective temperature of $T_\mathrm{eff} = 5638\pm 50\,$K, surface gravity of $\log g = 4.42 \pm 0.10$\,dex, metallicity of $\mathrm{[M/H]}=-0.081\pm 0.080$\,dex, and rotational broadening of $v\sin I_\star = 32.16\pm0.35\,\mathrm{km\,s}^{-1}$. The velocities are stable at the $200\, \mathrm{m\,s}^{-1}$ level, and the observed variability is consistent with that expected from spot modulation of a young star. The spectra and associated line profiles are further explored in Section~\ref{sec:fp} to rule out astrophysical false positive scenarios where a nearby star hosting an eclipsing binary system may mimic the planetary transit signal. 

\begin{table}
	\caption{Reconnaissance radial velocities}
\begin{tabular}{rrr}
\toprule
BJD & Relative RV [$\mathrm{m\,s}^{-1}$] & RV Error  [$\mathrm{m\,s}^{-1}$] \\
\midrule
2460285.976102  &     0$^*$   &  134 \\
2460290.032236  &  -172   &  134 \\
2460290.828697  &  -275   &  121 \\
2460291.937212  &  -436   &  138 \\
2460300.795295  &  -600   &  113 \\
2460310.890082  &  -360   &  134 \\
\bottomrule
\end{tabular}\\
 $^*$\textit{Velocities reported relative to first observation}\label{tab:tresrv}
\end{table}

\section{Stellar Characterization}\label{sec:star}
\begin{table}
	\caption{Properties of \thisstar.}
	\label{tab:star}
	\begin{tabular}{lcc} 
            \toprule
		\,\,\,Parameter & Value & Source \\
		\midrule
  \textbf{Identifier}\\
            ~~~TIC ID & TIC 434398831 & 1\\
             ~~~2MASS & J06155022+1601261 & 2\\
             ~~~APASS &  43186912 & 3\\
             ~~~Gaia DR2 & 3369841304767966976 & 4\\
             ~~~Tycho-2& TYC 1314-01161-1 & 5\\
             ~~~UCAC4& 531-023839 & 6\\
             ~~~WISE & J061550.22+160125.9 & 7\\
		\textbf{Astrometry}\\
             ~~~Right Ascension (RA) \dotfill &  06:15:50.22 & 8\\
             ~~~Declination (Dec) \dotfill & +16:01:25.85 & 8\\
             ~~~Parallax (mas) \dotfill & 4.643$\pm$0.017& 8\\
             \textbf{Proper Motion}\\
             ~~~RA Proper Motion (mas yr$^{-1}$) \dotfill & $5.528\pm0.072$ & 8\\
             ~~~Dec Proper Motion (mas yr$^{-1}$) \dotfill &$-16.428\pm0.061$ & 8\\
             \textbf{Photometry}\\
             ~~~\tess\ (mag) \dotfill &$11.311\pm0.011$ & 1\\
             ~~~\emph{B} (mag) \dotfill & $12.48\pm0.38$& 3\\
             ~~~\emph{V} (mag) \dotfill &$11.682 \pm 0.027$ & 3\\
             ~~~\emph{J} (mag) \dotfill & $10.590 \pm 0.021$ & 2\\
             ~~~\emph{H} (mag) \dotfill & $10.241 \pm 0.020$ & 2\\
             ~~~\emph{K} (mag) \dotfill & $
10.138 \pm 0.017$ & 2\\
             ~~~\emph{Gaia} (mag) \dotfill & $
11.7893 \pm 0.0038$ & 8\\
             ~~~$\mathrm{\emph{Gaia}}_\mathrm{BP}$ (mag) \dotfill &12.130$\pm0.013$& 8\\
             ~~~$\mathrm{\emph{Gaia}}_\mathrm{RP}$ (mag) \dotfill & 11.232$\pm0.010$ & 8\\
              ~~~WISE W1 (mag) \dotfill & 9.936$\pm$0.021
& 7\\
              ~~~WISE W2 (mag) \dotfill &10.048$\pm$0.019 & 7\\
              ~~~WISE W3 (mag) \dotfill &9.985$\pm$0.066 & 7\\
              ~~~WISE W4 (mag) \dotfill & 8.437& 7\\
              \textbf{Kinematics and Position}\\
              ~~~$U$ ($\mathrm{km\,s^{-1}}$) \dotfill &-18.26$\pm$1.50 & 8\\
              ~~~$V$ ($\mathrm{km\,s^{-1}}$) \dotfill & -22.73$\pm$0.39& 8\\
              ~~~$W$ ($\mathrm{km\,s^{-1}}$) \dotfill & -3.282$\pm$0.022& 8\\
              ~~~Distance (pc) \dotfill & $214.1_{-1.7}^{+1.7}$& 9\\
              \textbf{Physical Properties}\\
              ~~~$M_\star$ (M$_\odot$) \dotfill & $0.992_{-0.018}^{+0.014}$& 9\\[0.095cm]
              ~~~$R_\star$ (R$_\odot$) \dotfill & $0.909_{-0.048}^{+0.048}$& 9\\[0.045cm]
              ~~~$T_\mathrm{eff}$ (K) \dotfill & $5638 \pm 50$ & 9\\
              ~~~Surface gravity $\log g_\star$ (cgs) \dotfill & $4.42 \pm 0.10$& 9\\
              ~~~[m/H] \dotfill & $-0.081 \pm 0.080$ & 9\\
              ~~~$v\sin I_\star$ ($\mathrm{km\,s^{-1}}$) \dotfill & $32.16\pm0.35$ & 9\\
              ~~~Age (Myr) \dotfill & $61\pm6$ Myr & 9\\
              \textbf{Limb darkening coefficients}\\
              ~~~$u_{1,\mathrm{\tess}}$ \dotfill  & 0.349$\pm$0.026& 10\\
              ~~~$u_{2,\mathrm{\tess}}$ \dotfill  & 0.256$\pm$0.014& 10\\
              ~~~$u_{1,\mathrm{\emph{CHEOPS}}}$ \dotfill  & 0.467$\pm$0.036& 10\\
              ~~~$u_{2,\mathrm{\emph{CHEOPS}}}$ \dotfill  & 0.215$\pm$0.027& 10\\
              ~~~$u_{1,\mathrm{\emph{i'}}}$ \dotfill  & 0.364$\pm$0.030& 10\\
              ~~~$u_{2,\mathrm{\emph{i'}}}$ \dotfill  & 0.238$\pm$0.020& 10\\
              ~~~$u_{1,\mathrm{\emph{g'}}}$ \dotfill  & 0.656$\pm$0.049& 10\\
              ~~~$u_{2,\mathrm{\emph{g'}}}$ \dotfill  & 0.139$\pm$0.045& 10\\
              ~~~$u_{1,\mathrm{\emph{r'}}}$ \dotfill  & 0.460$\pm$0.039& 10\\
              ~~~$u_{2,\mathrm{\emph{r'}}}$ \dotfill  & 0.229$\pm$0.030& 10\\
              ~~~$u_{1,\mathrm{\emph{zs}}}$ \dotfill  & 0.285$\pm$0.032& 10\\
              ~~~$u_{2,\mathrm{\emph{zs}}}$ \dotfill  & 0.251$\pm$0.031& 10\\
              ~~~$u_{1,\mathrm{\emph{R}}}$ \dotfill  & 0.434$\pm$0.036& 10\\
              ~~~$u_{2,\mathrm{\emph{R}}}$ \dotfill  & 0.233$\pm$0.026& 10\\
              \textbf{Activity indicators}\\
               ~~~$P_{\mathrm{rot}}$ (days) \dotfill &  1.75$\pm0.12$& 9\\
               ~~~Li $6708\,$\AA\, EW (\AA) \dotfill & 0.205$\pm$0.089 & 9\\
		\bottomrule
	\end{tabular}
 $^1$\cite{Stassun:2019}; $^2$\cite{2MASS}; $^3$\cite{APASS}; $^4$\cite{GaiaDR22018}; $^5$\cite{Hog:2000}; $^6$\cite{UCAC}; $^7$\cite{WISE}; $^8$\cite{GaiaDR3}; $^9$This work; $^{10}$Interpolated from \cite{Claret2017}.
\end{table}

\subsection{\thisstar\ and Theia 116}

\begin{figure}
    \centering
    \includegraphics[width=\linewidth]{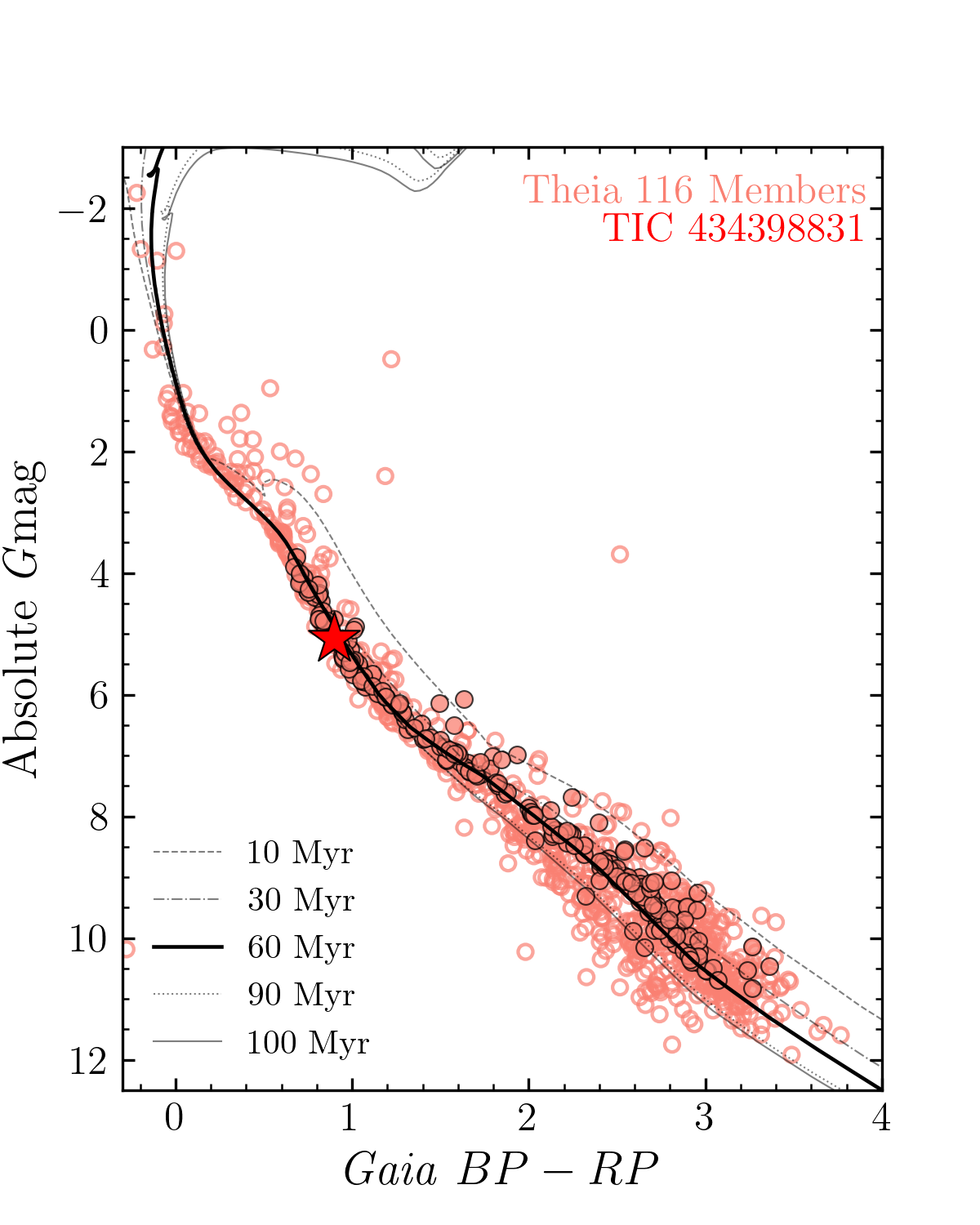}
    \caption{Color-Magnitude Diagram (CMD) for candidate Theia 116 members and \thisstar\ (red star). We plot candidate members with measured rotation periods in orange, filled circles. Candidates without measured rotations are plotted as empty circles.}
    \label{fig:cmd}
\end{figure}

\thisstar\ is a kinematic member of the Theia 116 association \citep{Kounkel:2019}. We detail below our own independent analysis of the properties of the Theia 116 association, including isochrone fitting of its members, \tess\ derived rotation distribution, and a limited lithium survey of its members to determine the age of the group.

\subsubsection{Kinematics}

\begin{figure*}
    \centering
    \includegraphics[width=0.95\linewidth]{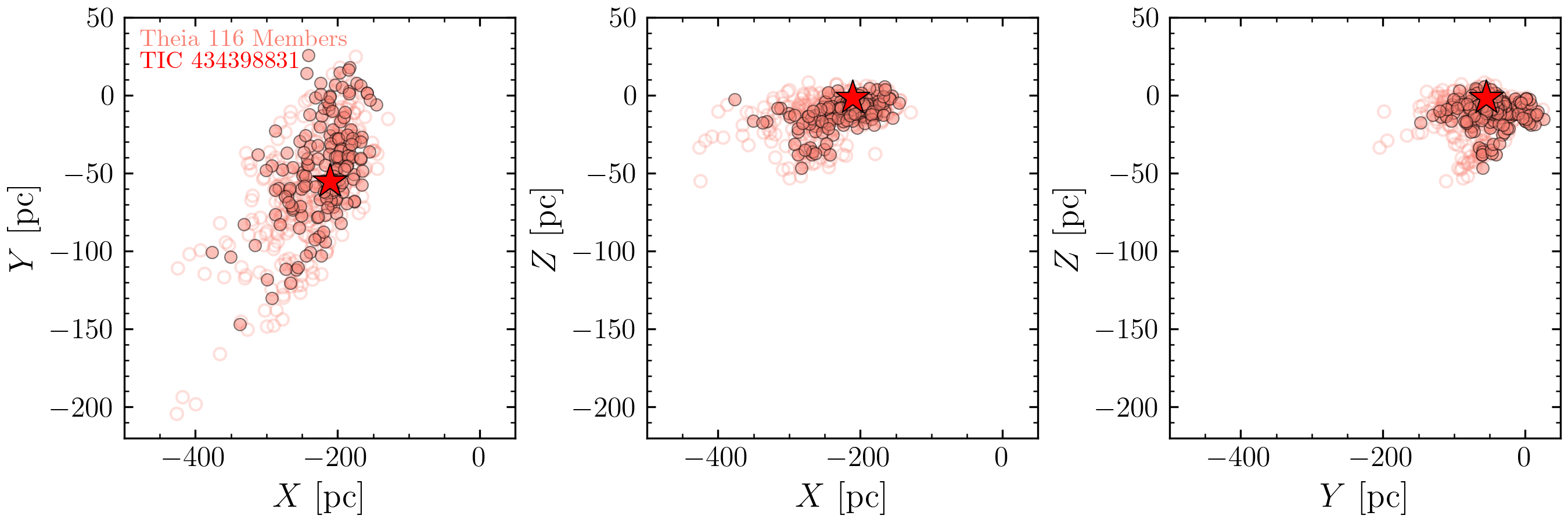}\\
    \includegraphics[width=0.95\linewidth]{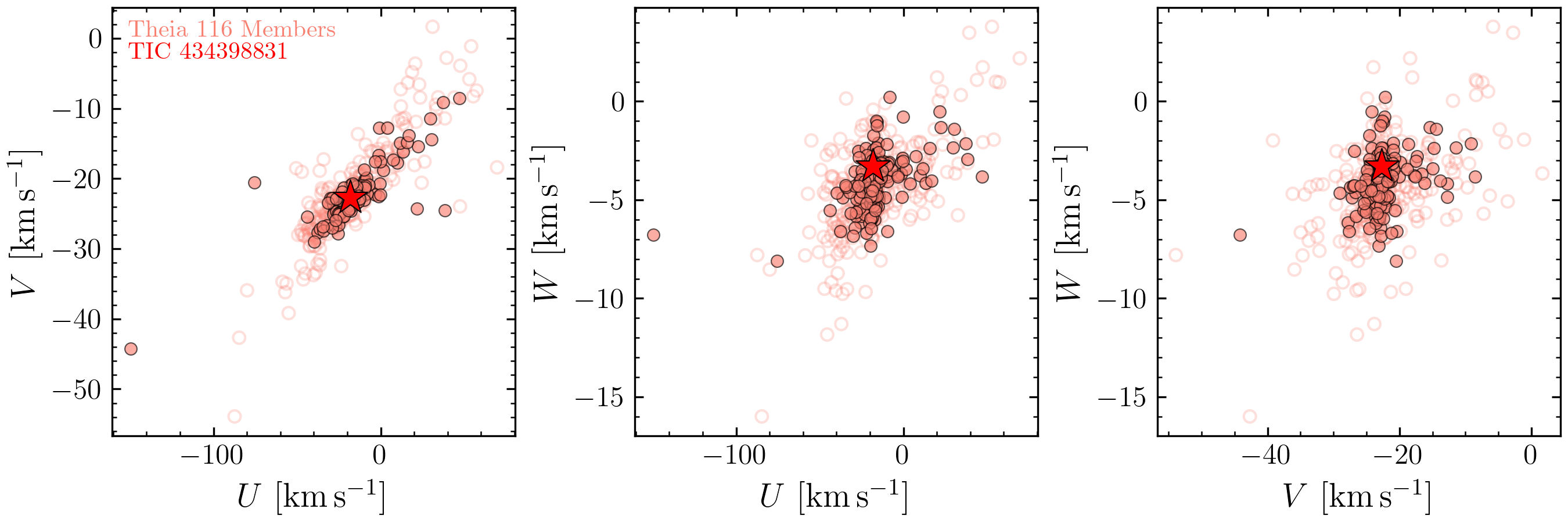}
    \caption{Galactic $X,Y,Z$ positions (top panel) and $U,V,W$ spatial velocities (bottom panel) of \thisstar\ (red star) and candidate Theia 116 members (circles).  Candidate members with detected rotations from \tess\ FFI light curves are filled in.}
    \label{fig:xyzuvw}
\end{figure*}

\thisstar\ was identified as a candidate member of Theia 116 in \citet{Kounkel:2019}. In brief, \citet{Kounkel:2019} performed a search for comoving stellar populations through analysis of spatial motions from \textit{Gaia} DR2 \citep[][]{GaiaDR2}{}{} via the unsupervised learning algorithm HDBSCAN \citep[][]{McInnes:2017}{}{}. The ages of identified stellar groups were estimated via fitting an isochrone to the groups' color-magnitude diagram (see Figure~\ref{fig:cmd}). 
\citet{Kounkel:2019} identified 1108 candidate members of Theia 116. We used \textit{Gaia} DR3 \citep[][]{GaiaDR3}{}{} to determine Theia 116 candidate members' spatial positions and velocities. Figure~\ref{fig:xyzuvw} shows \thisstar\ relative to the space motion distribution of the Theia 116 group. 

\subsubsection{Stellar Rotation}\label{sec:rot}

\begin{figure}
    \centering
    \includegraphics[width=\linewidth]{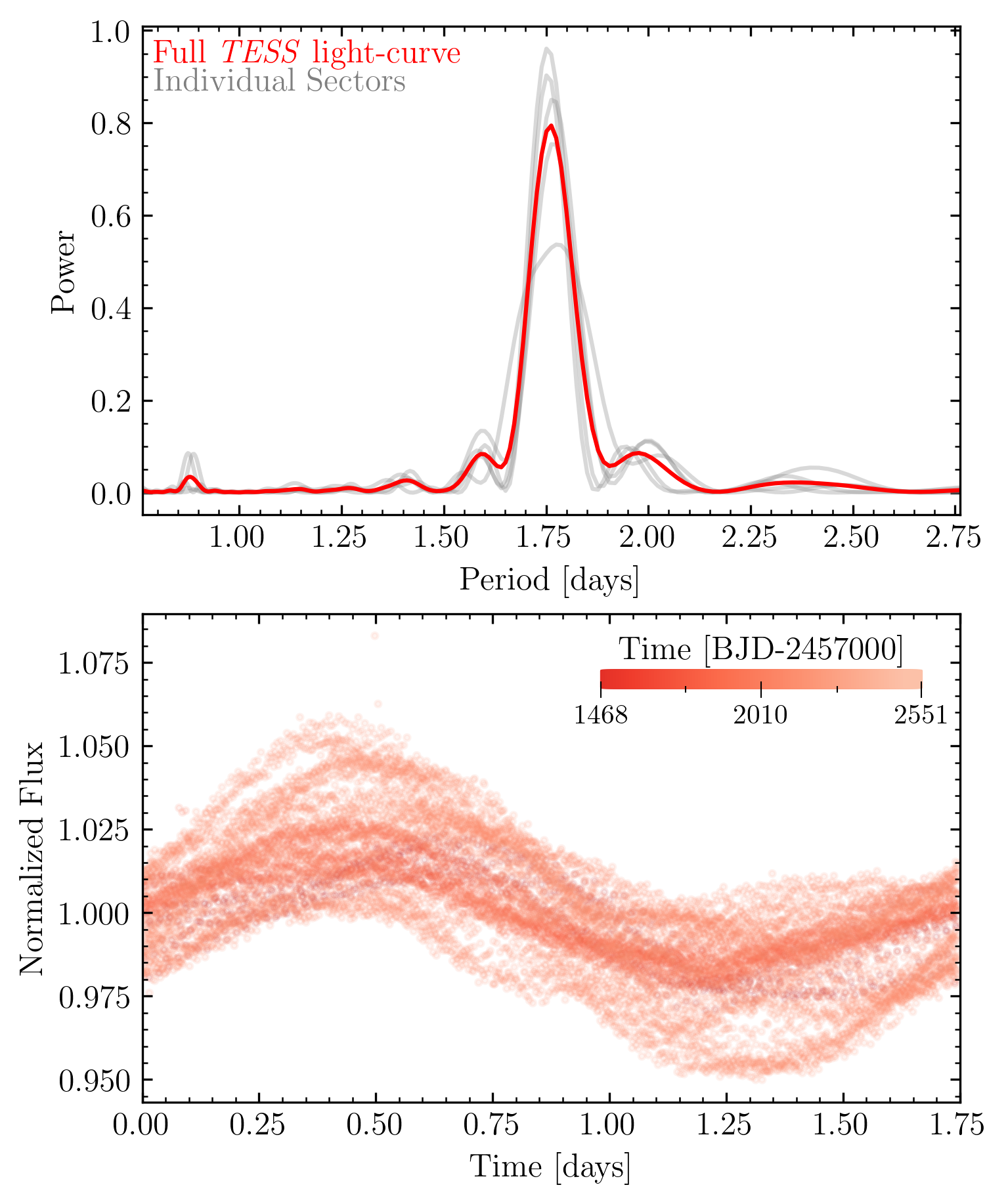}
    \caption{\textit{Top panel}: Lomb-Scargle periodogram for \thisstar\ both the average of the full \tess\ light curve (red) and each sector of \tess\ observations (grey). We measure a stellar rotation period of $P_\mathrm{rot} = 1.75\pm0.12$ days. \\
    \textit{Bottom panel}: Phase folded \tess\ light curve. The phased light curve is colored based on the time observation.}
    \label{fig:periodogram}
\end{figure}
\begin{figure*}
    \centering
    \includegraphics[width=0.95\linewidth]{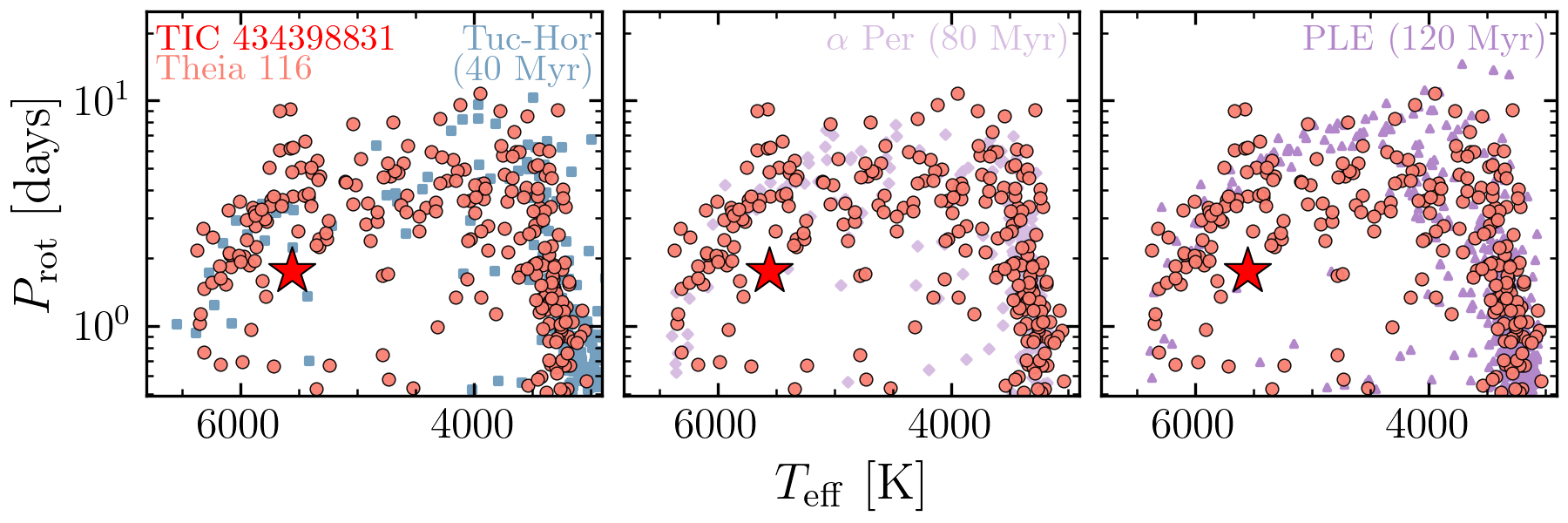}\\
    \includegraphics[width=0.95\linewidth]{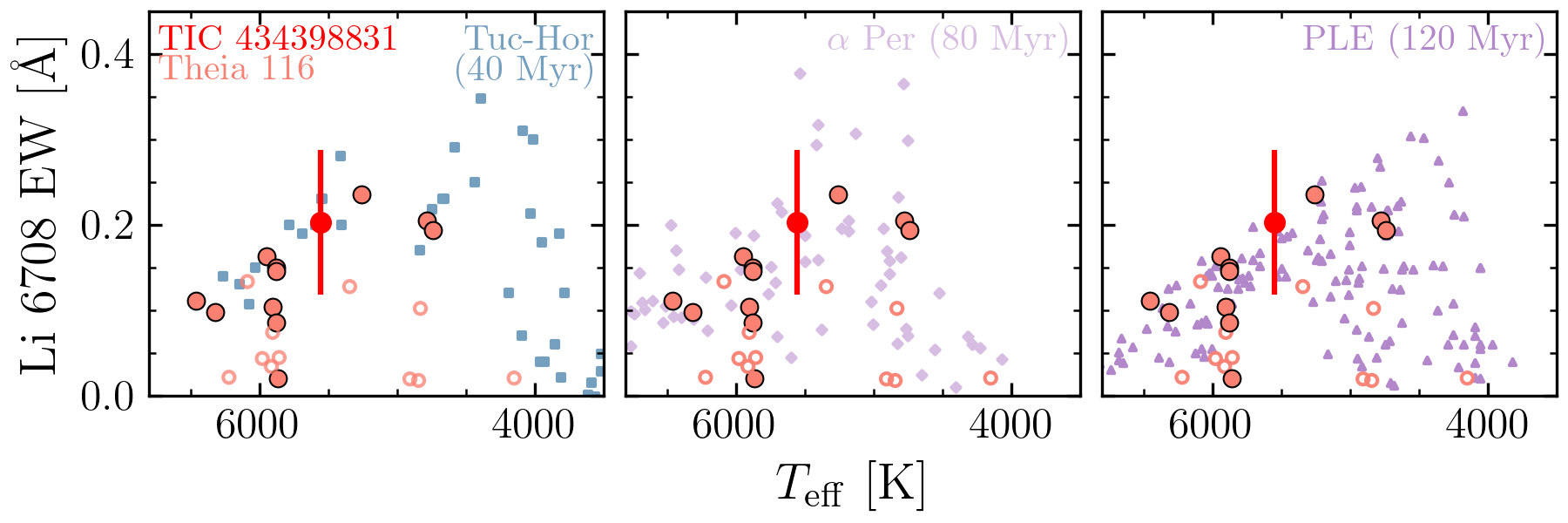}
    \caption{\textit{Upper panel}: Rotation periods for rotationally validated Theia 116 members (orange circles) and \thisstar\ (red star) as a function of stellar effective temperature. We plot measured rotations of well-characterized clusters for comparison-- 40 Myr Tuc-Hor \citep[left,][]{Popinchalk:2023}{}{}, 80 Myr $\alpha$\,Per \citep[middle,][]{Boyle:2023}{}{}, and 120 Myr Pleiades \citep[right,][]{Rebull:2016a}{}{}. \textit{Lower panel}: Lithium 6708\AA\, doublet equivalent widths (EW) of a subsample of Theia 116 members (orange circles) and \thisstar\ (red circle). Rotationally validated members of Theia 116 are the filled circles, while empty circles do not exhibit photometric signatures of youth, and are likely contaminating field stars. Literature Li EW's of Tuc-Hor \citep[][]{Popinchalk:2023}{}{}, $\alpha$\,Per \citep[][]{Balachandran:2011}{}{}, and Pleiades \citep[][]{Bouvier:2018}{}{} members are plotted for comparison. The spread in both rotation and Li abundances of Theia 116 members are consistent with youth. The rotation spread is more consistent with spreads of the pre-zero age main sequence (ZAMS) clusters Tuc-Hor and $\alpha$\,Per, than the spread of the ZAMS Pleiades cluster. }
    \label{fig:prot}
\end{figure*}
We make use of \tess\ Cluster Difference Imaging Photometric Survey \citep[CDIPS;][]{CDIPS}{}{} light curves for Theia 116 members to measure their stellar rotation periods. Groups of coevolved stars are born with a broad distribution of stellar rotation periods. As they contract onto the main sequence, the rotation distribution falls onto a gyrochrone, the distribution of the rotation periods of a set of coevolved stars as a function of stellar mass and age. For pre-main sequence stars, a spread in their rotation period distribution is expected as contraction occurs at different rates for different stars depending upon their stellar masses. Placing exact age constraints on associations with pre-main sequence stars is therefore difficult, but the rotation period distribution of a population can still be used to gauge the youth of the association.  

We cross-matched Theia 116 candidate members with the \tess\ Input Catalog \citep[TICv8; ][]{Stassun:2019}{}{}, and queried MAST for CDIPS light curves for member stars. We filtered for stars with stellar effective temperatures, $T_\mathrm{eff}$, between 3800 and 6200 K, as \texttt{gyro-interp} \citep[][]{bouma2023}{}{} is calibrated for stars in this temperature range, and a \textit{RUWE}$<1.4$ to remove potential binaries. This yields 701 stars with CDIPS light curves. We applied a Lomb-Scargle \citep[][]{lomb1976, scargle1982}{}{} periodogram to the light curves. We searched for periods between 0.5 and 12 days, due to challenges in deriving a rotation period longer than the duration of a \tess\ orbit ($\mysim13$ days).  

We visually examined the full light curve of each candidate member for (1) rotational variability, (2) binarity, (3) non-rotation variability, or (4) no variability. We also checked for aliasing of the true rotation period by phase-folding the light curve at the measured Lomb-Scargle period, half, and double the period. Candidate members are considered rotationally validated if they were classified as rotationally variable stars and we were able to confidently measure a stellar rotation period from the CDIPS light curve. We were able to rotationally validate 224 members of Theia 116, including \thisstar, for which we measured a rotation period of $P_\mathrm{rot} = 1.75\pm0.12$ days (see Figure~\ref{fig:periodogram}). We note the beat-like pattern in the \thisstar\ light curve, corresponding to the additional two peaks surrounding the maximum peak within the periodogram (see Figure~\ref{fig:periodogram}). These peaks are within 20\% of the derived rotation period. While a secondary source could be responsible for the observed beat-like pattern \citep[e.g.,][]{Stauffer:2018}, the observed features in the light curve and corresponding peaks in the periodogram are fully consistent with differential rotation of the young star \citep[][]{Aigrain:2015, Rebull:2016b, Santos, Stauffer:2018}. 
Our measured rotation periods for \thisstar\ and Theia 116 members are shown in Figure~\ref{fig:prot}.

To derive the gyrochonological cluster age, the stellar rotation periods, $P_\mathrm{rot}$, of the rotationally validated members were modeled with a gyrochrone using \texttt{gyro-interp}. For this analysis, we adopted $T_\mathrm{eff}$ values from the TIC. We implemented an MCMC via \texttt{emcee} \citep[][]{emcee}{}{} to explore the best-fit cluster age and its associated uncertainties. We derived a gyrochonological age of 1-80 Myr for Theia 116. As members of Theia 116 are still contracting onto the main sequence, our \texttt{gyro-interp} age is only able to provide an upper-limit for the cluster age, $<80$ Myr.

\subsubsection{Lithium Depletion}\label{sec:li}
\begin{table}
	\caption{TRES spectroscopic observations of Theia 116 members.}
\begin{tabular}{lccc}
\toprule
TIC & & Li 6708\AA\, EW [\AA] & $T_\mathrm{eff}$ [K]\\
\midrule
\textbf{434398831} &   & 0.205$\pm$0.088 &  5550$\pm$50 \\
 20311523$^*$ &   & 0.133$\pm$0.086 & 6090$\pm$50 \\
 44936347$^*$ &   & 0.0340$\pm$0.0078 & 5910$\pm$50 \\
 59350234$^*$ &   & 0.021$\pm$0.017 & 6220$\pm$50 \\
 75163937 &   & 0.098$\pm$0.224 & 6320$\pm$50 \\
 75950411 &   & 0.103$\pm$0.081 & 5910$\pm$50 \\
 76798674 &   & 0.085$\pm$0.022 & 5880$\pm$50 \\
 97633461$^*$ &   & 0.0739$\pm$0.0055 & 5900$\pm$50 \\
119036581 &   & 0.021$\pm$0.049 & 5860$\pm$50 \\
119039300 &   & 0.163$\pm$0.080 & 5950$\pm$50 \\
119191568$^*$ &   & 0.0445$\pm$0.0048 & 5860$\pm$50 \\
153214340$^*$ &   & 0.0434$\pm$0.0061 & 5980$\pm$50 \\
200527241 &   & 0.111$\pm$0.097 & 6460$\pm$50 \\
206723333 &   & 0.205$\pm$0.080&4780$\pm$50\\
247597942 &   & 0.1940$\pm$0.0078&4740$\pm$50\\
336889008 &   & 0.1020$\pm$0.0078&4830$\pm$50 \\
336984853 &   & 0.150$\pm$0.048 & 5880$\pm$50 \\
358497174 &   & 0.018$\pm$0.010&	4850$\pm$50 \\
386044087 &   & 0.235$\pm$0.079 & 5260$\pm$50 \\
415560146 &   & 0.0195$\pm$0.0087 & 4910$\pm$50 \\
429494528 &   & 0.146$\pm$0.072 & 5880$\pm$50 \\
434380619$^*$ &   & 0.13$\pm$0.28 & 5350$\pm$50 \\
438003891 &   & 0.021$\pm$0.010 & 4150$\pm$50	\\
\bottomrule
\end{tabular}\\
 $^*$\textit{No photometric signatures of youth in \tess\ light curves.}\label{tab:li}
\end{table}

We attempted to sample the atmospheric lithium abundances of the stellar population to measure the lithium age of Theia 116. As low-mass stars contract and their core temperatures surpass a threshold temperature, lithium burning is triggered \citep[][]{Soderblom:2014}{}{}. Convective envelopes transport the lithium from the stellar surface to the core, where it is burned, thereby decreasing the surface lithium abundance. This process occurs most rapidly at lower masses, with fully convective stars depleting their lithium on time scales of $\mysim10\,$Myr. For stellar groups, the distribution of lithium abundances as a function of stellar mass enables precise age measurements. 


We obtained single epoch spectra for 22 members of the Theia 116 association (16 rotationally validated members) using TRES. These members were randomly selected across different effective temperatures, ranging from 4000\,K to 6500\,K to best sample the lithium distribution. 

We measured the equivalent width of the lithium doublet at 6708\,\AA{} as per \citet{Zhou2021}. We first performed a least-squares deconvolution analysis to derive the line profile of each star, and thereby measure its rotational velocity. We then fit the spectral region around the 6708\,\AA{} line via a set of Gaussian profiles. We assumed the two lines of the lithium doublet have the same amplitude and the same width as that of the stellar rotational broadening velocity. We also simultaneously modeled the nearby Fe I line at 6707.43\,\AA{} to account for its contamination to the lithium doublet line strength, adopting the same broadening width, but allowing its amplitude to be free. We adopted the integral below our model of the lithium doublet as the equivalent width of the lithium line. All spectra were checked for signatures of binarity, no members exhibited double-line profiles. Effective temperatures were also derived from each spectrum via SPC as per Section~\ref{sec:spec}. The measured lithium abundances are presented in Table~\ref{tab:li}. We compare our measured lithium equivalent widths against literature values for well-characterized clusters in Figure~\ref{fig:prot}. 

We made use of the lithium depletion empherical model available in the `Empirical AGes from Lithium Equivalent widthS’ \citep[\texttt{EAGLES;}][ Eq. 4]{eagles} to derive the cluster's lithium abundance age. For members with both LiEW measurements and \tess\ CDIPS light curves, we visually inspected the CDIPS light curves for signatures of youth. We excluded LiEW measurements for previously identified members that show no signatures of photometric variability consistent with youth, as they are likely contaminating field stars. We derived the best-fit lithium abundance age and explored its posteriors using \texttt{emcee}. We measured a lithium abundance age of $68\pm4$ Myr. 

\subsubsection{Isochrone modeling of membership SEDs}

\begin{figure}
    \includegraphics[width=\linewidth]{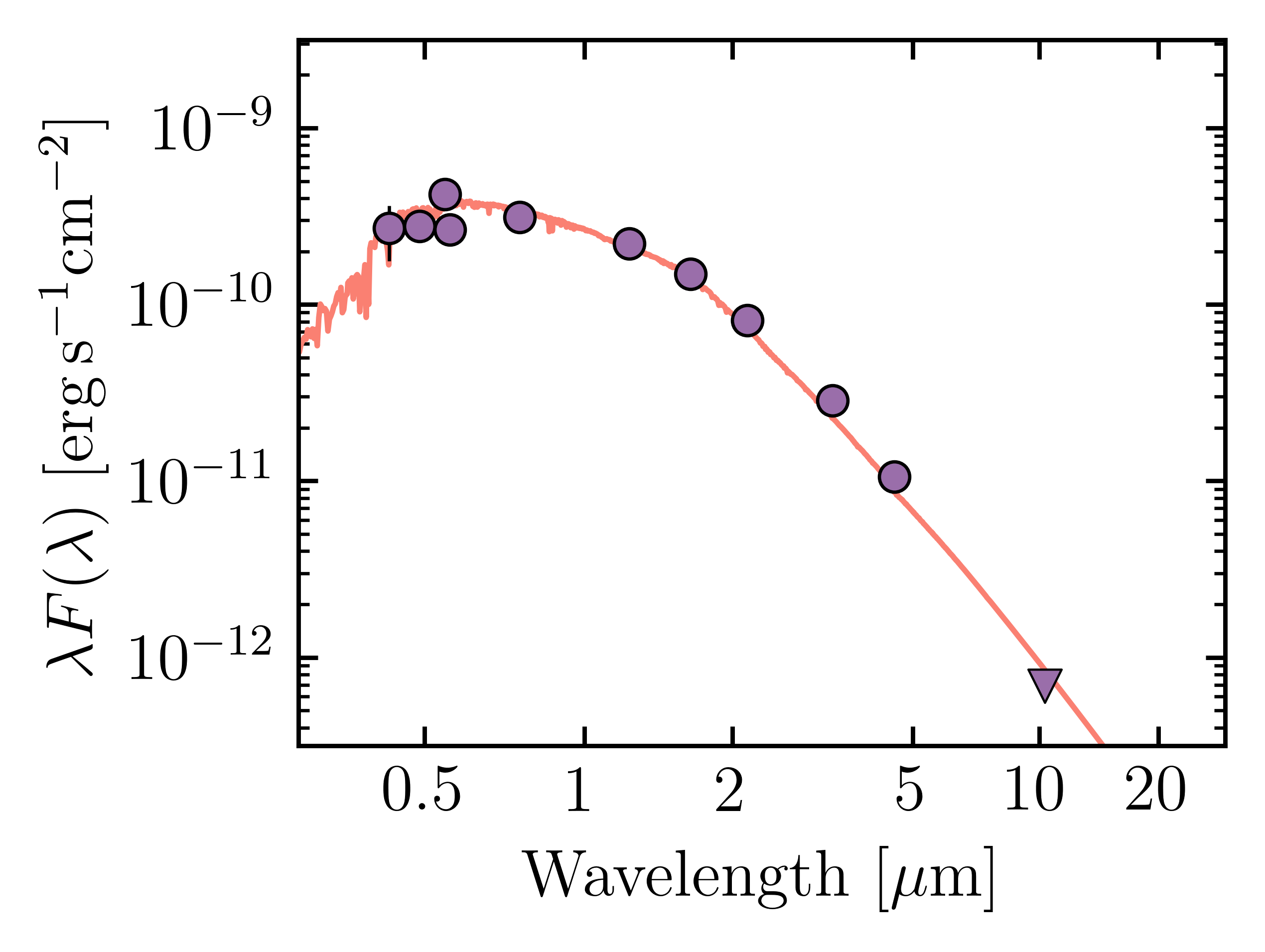}
    \caption{The SED of \thisstar\ from \textit{2MASS}, \textit{Gaia}, and \textit{WISE} photometry. The \textit{WISE} W4 magnitude (triangle) is an upper limit. }
    \label{fig:sed}
\end{figure}

We made use of available literature photometry for rotationally validated members of Theia 116 to derive a SED age. As with the analysis of the stellar rotations, we selected for members with \textit{Gaia} \textit{RUWE} values $<1.4$, in an attempt to filter for binaries \citep[][]{Kervella:2022}{}{}. Further, we required members to have observed magnitudes and uncertainties from \textit{Gaia} DR3 \citep[][]{GaiaDR3}{}{}, \textit{WISE} \citep[][]{WISE}{}{}, and \textit{2MASS} \citep{2MASS}.

We fitted for the cluster age, extinction, and stellar mass of each rotationally validated member.These parameters were then interpolated onto a MESA Isochrones and Stellar Tracks (MIST) isochrone \citep[][]{Dotter:2016}{}{} interpolated with the \texttt{minimint} code \citep[][]{Koposov:2021}{}{}. We adopt the absolute magnitudes that are provided by the MIST models and compute apparent magnitudes with the incorporation of the stars' known parallax and modeled extinction. We assumed solar metallicity and \textit{Gaia} parallax values. We compared estimated magnitudes to observed magnitudes from the \textit{Gaia} DR3 \textit{G}, \textit{BP}, and \textit{RP} bands, \textit{WISE} \textit{W1, W2,} and \textit{W3} bands, and \textit{2MASS} \textit{J, H,} and \textit{Ks} bands. The SED of \thisstar\ is presented in Figure~\ref{fig:sed}. The best-fit cluster age and its associated uncertainties were explored via an MCMC implemented in \texttt{emcee}. We estimated a SED cluster age of 55$\pm$4 Myr.

\subsubsection{Age of Theia 116}\label{sec:age}

\begin{figure}
\includegraphics[width=\linewidth]{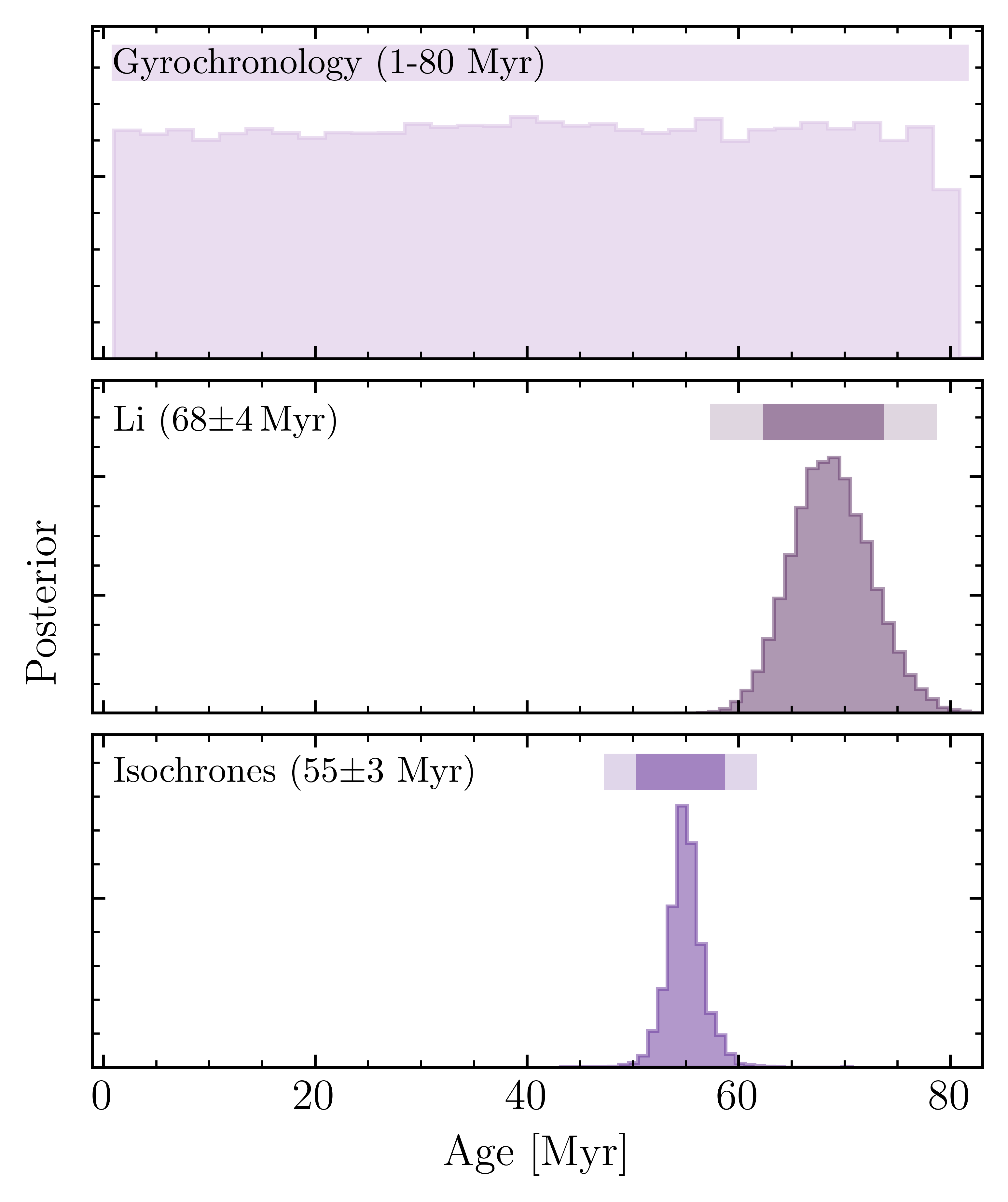}
    \caption{Theia 116 cluster age posteriors. We estimated the cluster age using gyrochronology, lithium abundances, and isochrone SED modeling independently. We adopt a conservative cluster age of \age.}
    \label{fig:age}
\end{figure}

While older than the initial isochronal age from \citet{Kounkel:2019}, $49_{-10}^{+13}$ Myr, our isochronal age estimate agrees with recent literature age estimates. \citet{Kounkel:2020} provided an updated age of $55_{-13}^{+17}$ Myr using the neural network \texttt{Auriga} \citep[][]{Auriga}{}{}. \citet{Barber:2023} derived a cluster age of $61_{-9}^{+13}$ Myr from excess \textit{Gaia} photometric variability. 

We attempted numerous joint age analyses that incorporated lithium distributions, isochrones, and rotation distributions together for Theia 116, but derived age posteriors with uncertainties too small to be realistic ($<3$ Myr), likely resulting from intrinsic uncertainties in stellar models unaccounted for in our analysis. For the remainder of this work, we therefore adopt a more conservative cluster age of 61$\pm6$ Myr, which encompasses the posteriors of the age analyses from independent lithium distribution modeling and the isochrone modeling. We present our age posteriors from the stellar rotation, lithium abundance, and SED analyses in Figure~\ref{fig:age}.


\subsection{Stellar flare rate of \thisstar}

In addition to heightened photometric variability induced by rapid rotation, stellar flares from active young stars bring additional challenges when searching for transit planets \citep[e.g. AU Mic c][]{Plavchan2020}{}{}. At $\mysim61$\,Myr, young G-type stars are still expected to have a heightened stellar flare rate when compared to their gigayear old counterparts \citep[][]{Feinstein:2024}{}{}. Visual examination of the \tess\ light curves showed evidence for stellar flares detected in the 10 min FFI observations. However, most flare identification algorithms are tuned for \tess\ 20 second or 2 minute cadence data \citep[e.g.][]{feinstein2020}{}{}. 

Following \cite{Howard:2022}, we identified flares as deviations $\geq4.5\sigma$ above the detrended \tess\ light curve. From \tess\ 30-minute and 10-minute observations in Sectors 6, 33, 43, 44, and 45, we identified 9 possible flares. These possible events were visually vetted, excluding detected deviations that occur at the edge of the \tess\ orbit. This resulted in 8 flare events across the 5 sectors of \tess\ observations. We derived a stellar flare rate of $2.0\times10^{-2}$ flares per day for \thisstar, slightly lower than expected stellar flare rates ($\mysim10^{-1}$ flares per day) for $>50$\,Myr G-type stars \citep[][]{Feinstein:2024}{}{}. It is possible the actual flare rate is higher, as stellar flares can occur on time scales shorter than 30 and/or 10 minutes. We repeated this exercise using the 200\,s data from sectors 71 and 72 which are likely to be more sensitive to flare events due to the shorter cadence \citep{Howard:2022b}. We identified 4 flare events (5 excursions from the detrended \tess\ light curve, 1 excluded event due to its location near the edge of the \tess\ orbit), and measured a flare rate of 1.01$\times10^{-1}$ flares per day in the 200 s data.

We repeated our flare rate calculation using the flare identification criteria implemented in \texttt{stella} \citep[][]{stella2020}, a convoluted neural network trained to detect flare events in the short cadence \tess\ data, to identify possible flare events in the 200\,s \tess\ observations from Sectors 71 and 72. \texttt{stella} requires candidate flare events to have an amplitude $\geq1.5\sigma$ above the detrended light curve, the two cadences immediately following the candidate event to have a $\geq1\sigma$ deviation above the detrended light curve as well, the cadences immediately before and after the candidate event must have lower amplitudes, and the amplitude of the candidate flare must be $\geq0.5\%$. Three candidate flare events within the 200s data meet this criterion. We note that due to the long cadence data, we were unable to use \texttt{stella} directly on the \tess\ data for \thisstar. This gives an estimated flare rate of 7.6$\times10^{-2}$ flares per day, which aligns with measured flare rates for young G dwarf stars with \texttt{stella} as per \citet[][]{Feinstein:2024}{}{}. The true flare rate may still be higher, as the apparent flare maximum can be reduced in the longer \tess\ cadences.

\section{Global Modeling}\label{sec:global_model}

\begin{table}
	\caption{\thisstarb\ and c best-fit parameters.}
	\label{tab:global_model}
	\begin{tabular}{lcccc} 
            \toprule
		\,\,\,Parameter & & Value & & Prior \\
		\midrule
		\textbf{TIC 434398831 b} & & & & \\[0.01cm]
            \,\,\, $T_0$ (BJD) \dotfill & &$2458468.6350_{-0.0034}^{+0.0026}$ & & Uniform\\[0.095cm]
            \,\,\, $P_b$ (days) \dotfill& & $3.685504_{-0.000011}^{+0.000012}$ & & Uniform \\[0.095cm]
             \,\,\, $R_{\mathrm{\emph{P,b}}}/R_\star$ \dotfill & &  $0.0353_{-0.0013}^{+0.0013}$ & &  Uniform\\ [0.095cm]
             \,\,\, $i$ (degrees) \dotfill & &  $88.2_{-1.1}^{+1.1}$ & &  Uniform\\[0.095cm]
             \,\,\, $\sqrt{e}\cos{\omega}$ \dotfill & & $0.000^{+0.069}_{-0.069}$ & & Uniform \\[0.095cm]
             \,\,\, $\sqrt{e}\sin{\omega}$ \dotfill & & $-0.010_{-0.061}^{+0.069}$ & & Uniform \\[0.095cm]
             \,\,\, $R_{\mathrm{\emph{P,b}}}$ ($\mathrm{R_\oplus}$) \dotfill & & $3.51_{-0.21}^{+0.22}$ & & Derived \\[0.095cm]
             \,\,\, $a/R_\star$ \dotfill & & 10.10$\pm0.11$ & & Derived \\[0.095cm]
             \,\,\, $a$ (AU) \dotfill & & $0.04641\pm0.00017$& & Derived \\[0.095cm]
             \,\,\, \textit{e} \dotfill & & $0.040^{+0.088}_{-0.034}$ & & Derived \\[0.095cm]
             \,\,\, $\omega$ (degrees) \dotfill & & $-61_{-51}^{+110}$ & & Derived \\[0.095cm]
           \,\,\, $T_\mathrm{eq}$ (K)$^*$ \dotfill & & $1250\pm11$ & & Derived \\[0.095cm]
            \,\,\, $T_{14}$ (hours) \dotfill & &$2.76\pm0.15$ & & Derived \\[0.095cm]
              \,\,\, $T_{23}$ (hours) \dotfill & &$2.54\pm0.16$ & & Derived \\[0.095cm]
            \textbf{TIC 434398831 c} & & & & \\
            \,\,\, $T_0$ (BJD) \dotfill & &$2458470.6239_{-0.0023}^{+0.0036}$ & & Uniform\\[0.095cm]
            \,\,\, $P_c$ (days) \dotfill& & $6.210291_{-0.000020}^{+0.000013}$ & & Uniform \\[0.095cm]
             \,\,\, $R_{\mathrm{\emph{P,c}}}/R_\star$ \dotfill & & $0.0568_{-0.0014}^{+0.0015}$ & & Uniform\\ [0.095cm]
             \,\,\, $i$ (degrees) \dotfill & & $87.79_{-0.61}^{+0.77}$ & & Uniform\\[0.095cm]
             \,\,\, $\sqrt{e}\cos{\omega}$ \dotfill & & $0.001_{-0.070}^{+0.069}$ & & Uniform \\[0.095cm]
             \,\,\, $\sqrt{e}\sin{\omega}$ \dotfill & & $0.008_{-0.070}^{+0.063}$ & & Uniform \\[0.095cm]
             \,\,\, $R_{\mathrm{\emph{P,c}}}$ ($\mathrm{R_\oplus}$) \dotfill & & $5.63_{-0.28}^{+0.29}$ & & Derived \\[0.095cm]
             \,\,\, $a/R_\star$ \dotfill & & $14.30^{+0.15}_{-0.16}$& & Derived \\[0.095cm]
             \,\,\, $a$ (AU) \dotfill & & $0.06572\pm0.00024$ & & Derived \\[0.095cm]
             \,\,\, \textit{e} \dotfill & & $0.013^{+0.025}_{-0.010}$& & Derived \\[0.095cm]
             \,\,\, $\omega$ (degrees) \dotfill & & $-59^{+106}_{-81}$& & Derived \\[0.095cm]
              \,\,\, $T_\mathrm{eq}$ (K)$^*$ \dotfill & &$1050\pm10$ & & Derived \\[0.095cm]
              \,\,\, $T_{14}$ (hours)\dotfill & &$3.04\pm0.26$ & & Derived \\[0.095cm]
              \,\,\, $T_{23}$ (hours) \dotfill & &$2.58\pm0.29$ & & Derived \\[0.095cm]
		\bottomrule
	\end{tabular}
 $^*$Assuming zero albedo, and planet retains all irradiation.
\end{table}

We performed a global model to derive the best-fit parameters for the \thisstar\ planetary system with all available \tess, \textit{CHEOPS}, ground-based photometry, and literature photometry. We jointly modeled planet transits and the stellar SED to best incorporate and propagate the associated uncertainties from each observation. 

Planet transits were modeled using the python implementation of the \citet{Mandel:2002} models known as \texttt{batman} \citep[][]{batman}{}{}. The free parameters in our global model describing the planet transits were the orbital period, $P$, time of transit center, $T_c$, radius ratio $R_P/R_\star$, orbital inclination, $i$ in degrees, and eccentricity terms, $\sqrt{e}\cos\omega$ and $\sqrt{e}\sin\omega$. We adopted quadratic limb-darkening parameters and fixed them at the values interpolated from \cite{Claret2017} based on the bandpasses of our photometric observations. 

For the \tess\ light curves, a 0.5\,day region around each transit was used for each transit modeling, with the rotational variability trend modeled via a third-order polynomial (see Figure~\ref{fig:tess_lc}). For the \textit{CHEOPS} and ground-based observations, we calculated and removed a trend model at each iteration. The light curve residuals, after removal of the tested transit model, were used to fit for a trend composed of a linear combination of the instrumental and environmental variables. These variables are listed in Section~\ref{sec:cheops} for the \textit{CHEOPS} light curve. For the ground-based light curves, these included the airmass, pixel positions $X$, $Y$, and full width at half maximum of the point spread function. This trend model was removed, and the log-likelihood was calculated from its residuals. 

The host star was described by our stellar free parameters, stellar mass, $M_\star$, stellar radius, $R_\star$, effective temperature, $T_\mathrm{eff}$, metallicity, [Fe/H], age, and parallax. The stellar parameters were interpolated onto a MIST isochrone \citep{Dotter:2016} with the python package \texttt{minimint} \citep{Koposov:2021} to estimate the stellar magnitudes. Derived stellar magnitudes were then compared against observed magnitudes from \textit{Gaia} DR3 \textit{G, BP,} and \textit{RP}, \textit{WISE} \textit{W1, W2,} and \textit{W3}, \textit{2MASS} \textit{J, H,} and \textit{Ks}, and \textit{APASS} \textit{B} and \textit{V} bands. 

We made use of our spectroscopic characterization of \thisstar\ (see Section~\ref{sec:spec}) to impose informed Gaussian priors on effective temperature, $T_\mathrm{eff}$, metallicity, [Fe/H], and surface gravity, $\log g_\star$. At each iteration, we require stellar parameters interpolated from the MIST isochrones to be within $3\sigma$ of the spectroscopic values.
We imposed Gaussian priors on the parallax from the measured \textit{Gaia} DR3 parallax and associated uncertainties, and stellar age from our cluster analysis (see Section~\ref{sec:age}). The best-fit values and their posteriors were explored through a Markov Chain Monte-Carlo with \texttt{emcee} \citep{emcee}. Our best-fit stellar and planetary parameters are presented in Tables~\ref{tab:star} and ~\ref{tab:global_model} respectively.

\section{Investigating possible false-positive scenarios}\label{sec:fp}

Due to the large size of \tess's pixels, some transit signals that pass vetting procedures may result from false positives. As \thisstarb\ and c were not released as a TOI, we investigate possible instrumental and astrophysical false-positive scenarios to validate the planets. The transit detections of both \thisstarb\ and c by multiple instruments ruled out instrumental artifact false-positive scenarios. Spectroscopic follow-up observations ruled out \thisstar being an eclipsing binary. 

LCO observations of \thisstarb\ and c detected the transits within a 3.1'' aperture. There are no \textit{Gaia} resolved stars within 3.1'' of \thisstar, with the nearest neighbor 12.31'' away. 
It is possible an unresolved, blended background star is responsible for the transit signals \citep[][]{Seager:2003}{}{}. We followed \citet{Vanderburg:2019} to estimate the brightest possible magnitudes that could induce the transit shapes observed:
\begin{equation}
    \Delta m_\mathrm{\textit{TESS}} \lesssim 2.5\log_{10} \left(\frac{t_{12}^2}{t_{13}^2\delta }\right)
\end{equation}
where $t_{12}$ is the ingress duration, $t_{13}$ is the time between first and third contact, and $\delta$ is the transit depth. An unresolved star would have to be within $\Delta m_\mathrm{\textit{TESS}} \lesssim 3.16$ and $\Delta m_\mathrm{\textit{TESS}} \lesssim 2.48$ (at a $3\sigma$ upper limit) to produce the signals corresponding to the transits of \thisstarb\ and c respectively. We estimated the density of stars within $\Delta m_\mathrm{\textit{TESS}} \leq 3.16$ in the general area of \thisstar\ to determine the likelihood of an unresolved background star being the source of the transit signals. We queried Gaia for resolved stars within a 1 arcmin cone centered around \thisstar, which yielded 32 stars. We cross-matched the Gaia sources with the TIC to calculate $\Delta m_{TESS}$. We derive a density of $4.42\times10^{-5}$ stars per square arcsecond within $\Delta m_\mathrm{\textit{TESS}} \leq 3.16$ of \thisstar.

\begin{figure}
    \centering
    \includegraphics[width=0.8\linewidth]{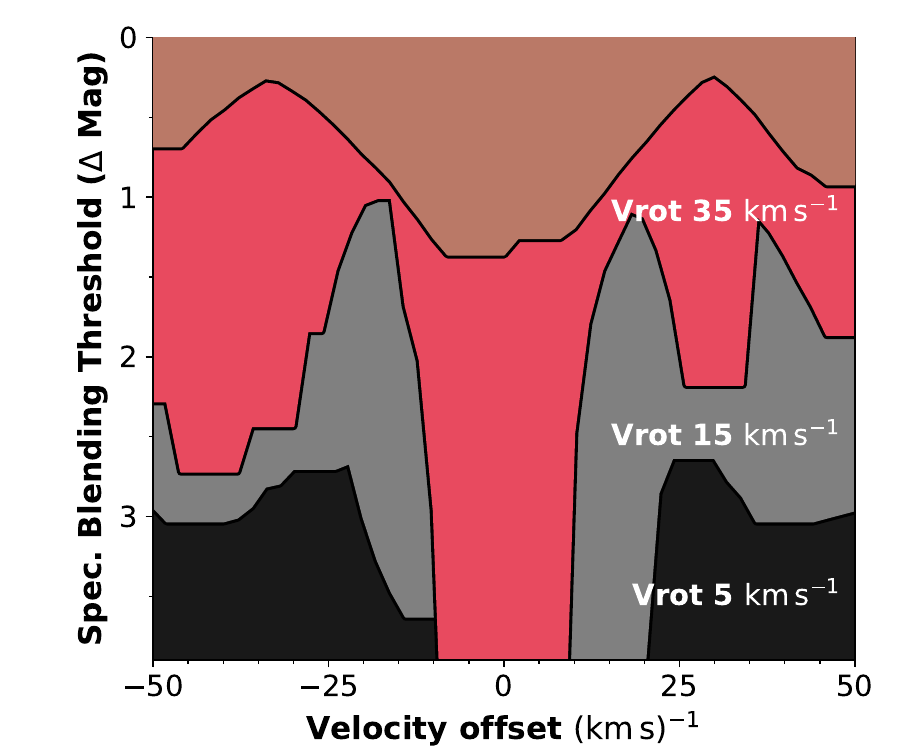}
    \caption{Detection thresholds for blend scenarios that can be rejected from spectroscopic line analysis. We can reject the presence of any slowly rotating companions ($v_\mathrm{rot} = 5\,\mathrm{km\,s}^{-1}$) within $\Delta m \lesssim3$ of \thisstar, shown in black. Similar constraints can be placed at weaker thresholds for $v_\mathrm{rot} = 15\,\mathrm{km\,s}^{-1}$ in grey and $v_\mathrm{rot} = 35\,\mathrm{km\,s}^{-1}$ in red.}
    \label{fig:spec_blend}
\end{figure}

We also made use of our TRES spectroscopic observations to help establish the magnitude limits of hypothetical stellar neighbors that may be blended with our target star. Following \citet{Zhou2021}, we used the line profiles derived from our TRES observations to determine if scenarios involving non-associated stellar companions with differing radial velocities can be ruled out, as they would yield a detectable set of lines in the spectra we observed. We injected a series of secondary signals into our average TRES line profile for \thisstar, varying in velocity separation (-50 to 50\,$\mathrm{km\,s}^{-1}$), line width (3 to 35\,$\mathrm{km\,s}^{-1}$), and primary-secondary flux ratio (0.001 to 0.2). We then attempted to recover each scenario with a double-lined model of the resulting line profiles. We found that for hypothetical sharp lined companions, $v_\mathrm{rot} = 5\,\mathrm{km\,s}^{-1}$, we can recover signals with $\Delta m \sim 3$ in the optical if the two stellar components are separated in velocity by $>10\mathrm{km\,s}^{-1}$. Recovery of stellar companions with more significant rotational broadening becomes progressively harder. Companions with rotational broadening similar to that of \thisstar\ can be recovered at the $\Delta m \sim 1$ level. Figure~\ref{fig:spec_blend} shows the magnitude limit of our spectroscopic blend analysis as a function of velocity separation from the host star. 

To further derive a false-positive probability for any remaining scenarios, we performed a joint analysis combining physically possible binary scenarios from \texttt{MOLUSC} \citep[][]{Wood:2021}{}{} with \texttt{TRICERATOPS} \citep[][]{Giacalone:2021}{}{}. We simulated 100,000 possible binary configurations between TIC434398831 and an unseen companion(s) with \texttt{MOLUSC} for both \thisstarb\ and c. We made use of radial velocity, HR-contrast curves, \textit{Gaia} RUWE and photometry to inform our \texttt{MOLUSC} simulations.

The simulated binaries from \texttt{MOLUSC} and our false-positive analyses were used to rule out \texttt{TRICERATOPS} false-positive scenarios (e.g., nearby eclipsing binaries, nearby transiting planets, dilution of the light curve due to an unresolved background star) that would not be physically realistic for \thisstar. As mentioned above, false-positive scenarios involving a resolved nearby star are ruled out through ground-based transit and HR-imaging observations. The remaining false-positive scenarios identified by \texttt{TRICERATOPS} were diluted transiting planets, resulting from an unresolved background star diluting the light curve of \thisstar. This would mean \thisstarb\ and c were not Neptune-sized planets but Jovian-sized. At the orbital periods of \thisstarb\ (\shortperb) and c (\shortperc), the occurrence rates for Jovian-sized planets are $<0.4\%$ \citep[][]{Kunimoto2020}{}{}. We therefore confidently present \thisstarb\ and c as validated planets, as the derived false-positive probabilities for \thisstarb\ and c are $0.0047_{-0.0013}^{+0.0012}$ and $0.000257_{-0.00022}^{+0.00556}$ respectively.

\section{Discussion and Conclusions}\label{sec:discussion}

We presented the discovery of \thisstar, a pre-main-sequence multiplanet system hosting a transiting mini- and super-Neptune. \thisstarb\ and c were originally identified in an independent survey searching for planets around stars in young comoving populations observed by \tess\ \citep[][]{Vach:2024}{}{}. Originally discovered in the \tess\ observations, \textit{CHEOPS} and LCO follow-up are able to constrain the orbital periods and transit times of \thisstarb\ and c. We perform a global model incorporating all available photometry to derive best-fit values for the system. We find \thisstarb\ is a mini-Neptune (\radb) at a \perb\ orbital period, and \thisstarc\ (\radc) is a super-Neptune at \perc. The ratios between the orbital periods are $\mysim1\%$ larger than the 5:3 mean motion resonance.  

\thisstar\ was identified to be part of the Theia 116 comoving population \citep[][]{Kounkel:2019}{}{}. Literature values estimate an age between 48-74 Myr for Theia 116. We independently derive a cluster age of \age\ using stellar rotation periods from \tess\ light curves, lithium equivalent width measurements, and SEDs of rotationally validated members. 

\begin{figure}
    \includegraphics[width=\linewidth]{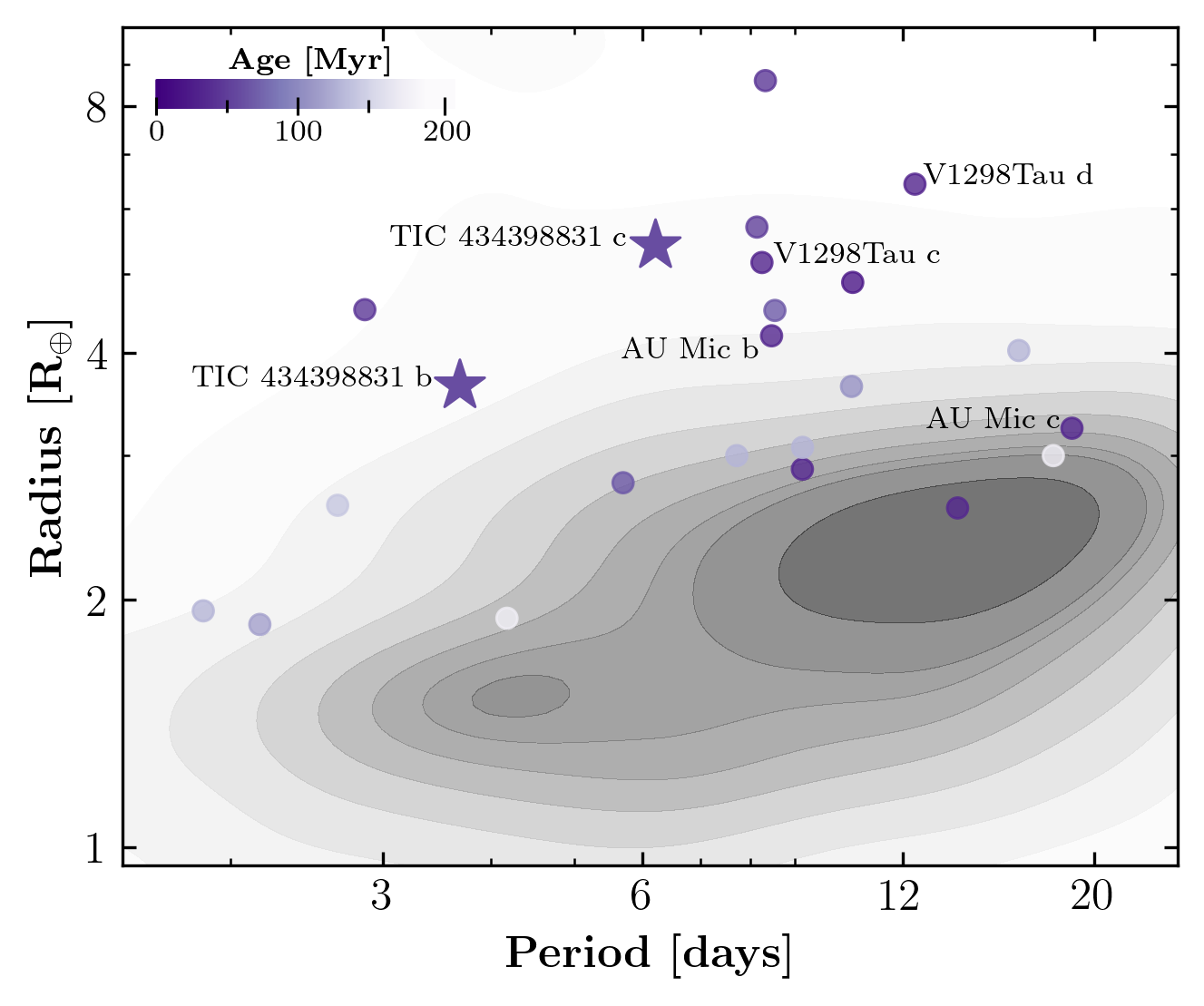}
    \caption{Radius-period distribution of the known young planet population. Planets are colored based on their ages, with darker purples representing the younger systems. \thisstarb\ and c are denoted by the stars. We plot the distribution of the known exoplanet population in greyscale for reference. Both \thisstarb\ and c lie above the mature exoplanet distribution. As they continue to evolve, both planets will likely evolve to lie in the \textit{Kepler} distribution. }
    \label{fig:population}
\end{figure}

The study of young planets allows us to test different mechanisms of planet formation and evolution against observations. Young multiplanet systems enable us to delve deeper into the impacts of orbital separation, as direct comparative planetology is enabled since the planets formed from the same disk and experienced the same stellar evolution. \thisstar\ joins a handful of multiplanet systems younger than 100 Myr which have proved invaluable in our understanding of young planets: V1298 Tau \citep{David:2019}, HD 109833 \citep{Wood:2023}, and AU Mic \citep{Plavchan2020}. \thisstar\ follows in trend with the young planet population, as both \thisstarb\ and c fall into a trough within the Kepler distribution (see Figure~\ref{fig:population}). We explore the predicted atmospheric evolution of \thisstarb\ and c in an attempt to understand how these young planets will compare to the Kepler distribution as they age.

\subsection{Atmospheric Evolution}

\begin{figure}
    \centering
    \includegraphics[width=\linewidth]{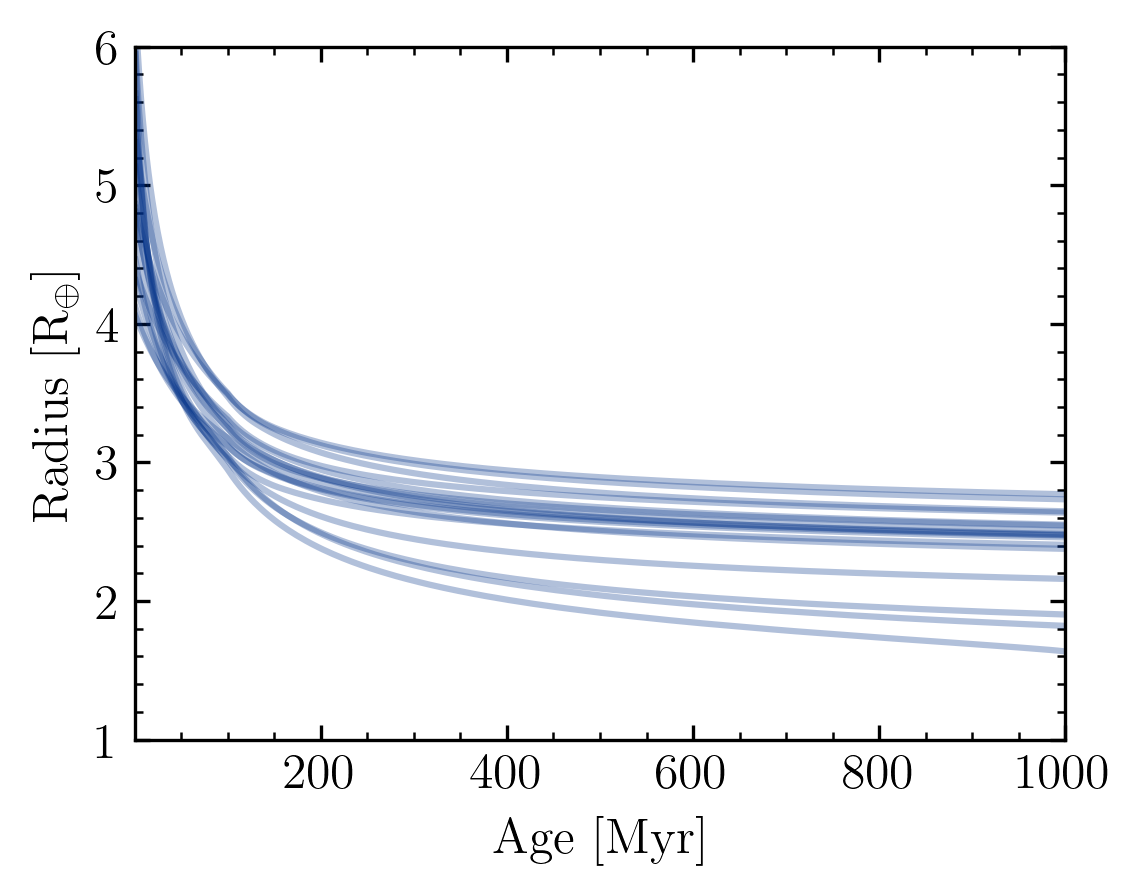}\\
    \includegraphics[width=\linewidth]{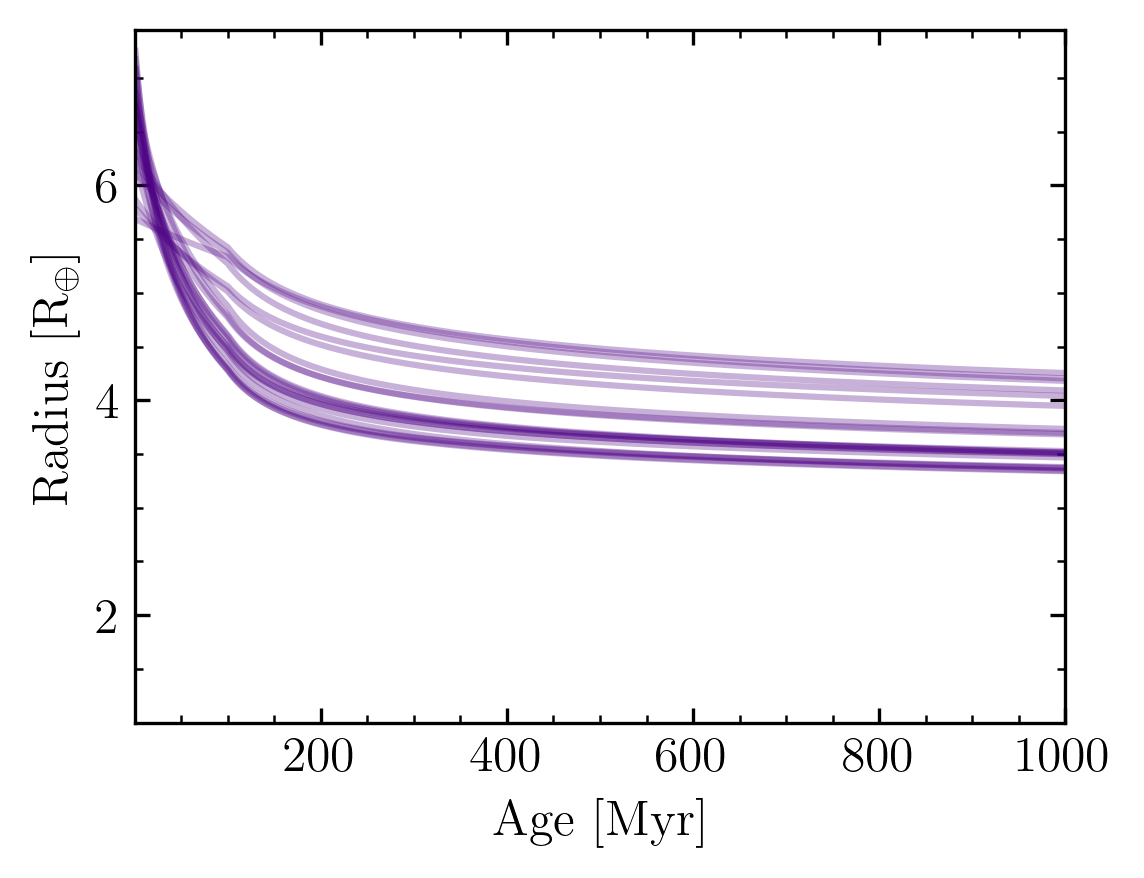}
    \caption{Sample of possible radius evolution tracks of \thisstarb\, (top) and c (bottom) undergoing thermal contraction and atmospheric erosion via photoevaporation escape \citep{owen:2013}. We present a sample of 30 MCMC chains to illustrate the various possible evolutionary outcomes for these planets. }
    \label{fig:radevol}
\end{figure}

Planets with H/He dominated atmospheres are thought to be born larger, and through evolutionary processes, experience contraction and mass loss resulting in a decrease in the observed radius. At $\mysim61$ Myr, \thisstarb\ and c are likely still experiencing mass loss via atmospheric erosion \citep[][]{owen:2013, Owen:2023, Gupta:2019, Rogers:2021, Rogers:2023}{}{}. We investigate the eventual fates of \thisstarb\ and c by simulating their radius evolution following the semi-analytic model for thermal and photoevaporative evolution from \citet{Owen:2017}.  We implement an MCMC using \texttt{emcee} to explore possible initial conditions and radius evolution of \thisstarb\ and c. We allow the core mass $M_{b,\mathrm{core}}$, bulk core density $\rho_{b,\mathrm{core}}$, initial envelope mass fraction $f_{env, b}$, and constant mass loss efficiency $\eta_b$ to be free parameters. The initial envelope mass fraction is defined at 10 Myr (i.e. disk dissipation time scale) as per literature standards. We fixed the stellar mass with our best-fit value derived in Section~\ref{sec:global_model} and adopted the empirical relation presented in \citet{Jackson:2012} to model the evolution of the high-energy luminosity of the host star. We make use of the MESA stellar evolution models interpolated with \texttt{minimint} using our best fit stellar mass and metallicity, at each step in our planetary evolution model, we adopt the corresponding stellar radius and effective temperature. We fixed the orbital period at our best-fit value, which is used to estimate the equilibrium temperature of the planet. We anchored the planet radius and the system age, rejecting chains that do not result in an evolutionary path that falls within the $1\sigma$ range of our measured planet radii at our system age and associated uncertainties. 

\thisstarb\ currently falls into the typical mini-Neptune regime ($2-4\,R_\oplus$), however, its eventual fate is unknown, as it may evolve across the radius valley to be a super-Earth ($1-2\,R_\oplus$). Our simulations predict $M_{b,\mathrm{core}} = 6.8_{-1.5}^{+1.9}\, M_\oplus$, $f_{env, b} = 0.41_{-0.21}^{+0.24}$, $\eta_c = 0.087_{-0.036}^{+0.073}$, and $\rho_{b,\mathrm{core}} = 4.5^{+1.3}_{-1.3}\, \mathrm{g\, cm^{-3}}$. While the majority of our simulated evolutionary paths predict \thisstarb\ to remain in the mini-Neptune regime, $\mysim2.5\%$ of our simulations result in \thisstarb\ being stripped of its volatile envelope and becoming a super-Earth. We present a sample of 30 evolutionary tracks for \thisstarb\ in Figure~\ref{fig:radevol}. 

\thisstarc\ is currently located in the super-Neptune valley, a region sparsely populated by the mature exoplanet population (see Figure~\ref{fig:population}). However, demographic studies of the young exoplanet population have unveiled an over-abundance of super-Neptune planets around the youngest stars \citep[>2.8$\sigma$;][]{Vach:2024}{}{}. The simulated evolutionary tracks predict $M_{c,\mathrm{core}} = 7.2_{-1.6}^{+1.8}\, M_\oplus$, $f_{env, c} = 0.65_{-0.15}^{+0.10}$, $\eta = 0.017_{-0.010}^{+0.021}$, and $\rho_{c,\mathrm{core}} = 4.4^{+1.5}_{-1.4}\, \mathrm{g\, cm^{-3}}$. The majority of the accepted evolutionary paths show \thisstarc\ evolving into the mini-Neptune regime, with the smallest predicted radii to be $\mysim3.4\,R_\oplus$. However, a non-negligible fraction of the accepted chains shows an evolutionary path resulting in a super-Neptune fate for \thisstarc. Within the mature exoplanet population, planets in this period-radius space are exceptionally rare, with an occurrence of 0.09 planets per 100 stars \citep[][]{Kunimoto2020}{}{}. 

Both planets are predicted to have significant initial mass envelope fractions, $f_{env, b} = 0.41$ and $f_{env, c} = 0.65$. These values are significantly larger than the predicted average initial mass envelope fraction for the \textit{Kepler} distribution, $\mysim0.10$. However, HST observations of the $\mysim23$\,Myr, V1298 Tau b revealed a H/He envelope fraction of $\mysim40\%$ \citep[][]{Barat:2024}{}{} similar to the predicted mass envelope fractions of \thisstarb\ and c.
While these simulations only account for mass-loss via photoevaporation, it is likely additional mass-loss mechanisms are at play and work in hand with photoevaporation to sculpt the distribution of the mature population \citep[e.g.][]{Gupta:2019,Owen:2023,Rogers:2024}{}{}.




\section*{Data Availability}

All \tess\ data products used in this paper are publically available through the Mikulski Archive for Space Telescopes (MAST). Ground-based observations used in this paper are available via ExoFOP-TESS. \textit{CHEOPS} data will be shared on reasonable request to the corresponding author.
 

\section*{Acknowledgements}
We respectfully acknowledge the traditional custodians of
the lands on which we conducted this research and throughout Australia. We recognize their continued cultural and spiritual connection to the land, waterways, cosmos, and community. We pay our deepest respects to all Elders, present
and emerging, and the people of the Giabal, Jarowair, and
Kambuwal nations, upon whose lands this research was conducted.

GZ thanks the support of the ARC DECRA program DE210101893 and ARC Future program FT230100517.
CH thanks the support of the ARC DECRA program DE200101840.

Funding for the \tess\ mission is provided by NASA's Science Mission directorate. This research has made use of the Exoplanet Follow-up Observation Program (EXOFOP) website, which is operated by the California Institute of Technology, under contract with the National Aeronautics and Space Administration under the Exoplanet Exploration Program. This paper includes data collected by the \tess\ mission, which are publicly available from the Mikulski Archive for Space Telescopes (MAST).

This work makes use of observations from the Las Cumbres Observatory global telescope network. 

This paper is based on observations made with the MuSCAT3 instrument, developed by the Astrobiology Center and under financial support by JSPS KAKENHI (JP18H05439) and JST PRESTO (JPMJPR1775), at Faulkes Telescope North on Maui, HI, operated by the Las Cumbres Observatory.
This work is partly supported by JSPS KAKENHI Grant Numbers JP24H00017 and JP24K00689, and JSPS Bilateral Program Number JPJSBP120249910.
The postdoctoral fellowship of KB is funded by F.R.S.-FNRS grant T.0109.20 and by the Francqui Foundation.
AS was supported by a NASA Massachusetts Space Grant Fellowship.
KKM acknowledges support from the New York Community Trust Fund for Astrophysical Research.

This work was enabled by observations made from the Gemini North telescope, located within the Maunakea Science Reserve and adjacent to the summit of Maunakea. We are grateful for the privilege of observing the Universe from a place that is unique in both its astronomical quality and its cultural significance. Observations in the paper made use of the High-Resolution Imaging instrument(s) Alopeke. Alopeke was funded by the NASA Exoplanet Exploration Program and built at the NASA Ames Research Center by Steve B. Howell, Nic Scott, Elliott P. Horch, and Emmett Quigley. Alopeke was mounted on the Gemini North telescope of the international Gemini Observatory, a program of NSF NOIRLab, which is managed by the Association of Universities for Research in Astronomy (AURA) under a cooperative agreement with the U.S. National Science Foundation. on behalf of the Gemini partnership: the U.S. National Science Foundation (United States), National Research Council (Canada), Agencia Nacional de Investigaci\'{o}n y Desarrollo (Chile), Ministerio de Ciencia, Tecnolog\'{i}a e Innovaci\'{o}n (Argentina), Minist\'{e}rio da Ci\^{e}ncia, Tecnologia, Inova\c{c}\~{o}es e Comunica\c{c}\~{o}es (Brazil), and Korea Astronomy and Space Science Institute (Republic of Korea). 


\bibliographystyle{mnras}
\bibliography{refs} 



\section*{Author Affiliations}
{
$^{1}$Centre for Astrophysics, University of Southern Queensland, West Street, Toowoomba, QLD 4350, Australia\\
$^2$Department of Physics and Astronomy, The University of North Carolina at Chapel Hill, Chapel Hill, NC 27599, USA\\
$^{3}$Center for Astrophysics \textbar{} Harvard \& Smithsonian, 60 Garden Street, Cambridge, MA 02138, USA\\
$^4$Department of Earth, Planetary, and Space Sciences, The University of California, Los Angeles, 595 Charles E. Young Drive East, Los Angeles, CA 90095, USA\\
$^{5}$Cahill Center for Astrophysics, California Institute of Technology, Pasadena, CA 91125, USA\\
$^{6}$Department of Physics, Lafayette College, 730 High St., Easton, PA 18042, USA\\
$^{7}$\MITKavli \\
$^{8}$George Mason University, 4400 University Drive, Fairfax, VA, 22030 USA\\
$^{9}$Kotizarovci Observatory, Sarsoni 90, 51216 Viskovo, Croatia\\
$^{10}$Astrobiology Research Unit, Universit\'e de Li\`ege, 19C All\'ee du 6 Ao\^ut, 4000 Li\`ege, Belgium\\
$^{11}$Department of Earth, Atmospheric and Planetary Sciences, Massachusetts Institute of Technology, 77 Massachusetts Ave, Cambridge, MA 02139, USA \\
$^{12}$Instituto de Astrof\'isica de Canarias (IAC), Calle V\'ia L\'actea s/n, 38200, La Laguna, Tenerife, Spain\\
$^{13}$Department of Physics and Astronomy, Wellesley College, Wellesley, MA 02481, USA\\
$^{14}$Komaba Institute for Science, The University of Tokyo, 3-8-1 Komaba, Meguro, Tokyo 153-8902, Japan\\
$^{15}$Astrobiology Center, 2-21-1 Osawa, Mitaka, Tokyo 181-8588, Japan\\
$^{16}$NASA Ames Research Center, Moffett Field, CA 94035, USA\\
$^{17}$Bay Area Environmental Research Institute, Moffett Field, CA 94035, USA\\
$^{18}$South African Astronomical Observatory, P.O. Box 9, Observatory, Cape Town 7935, South Africa\\
$^{19}$Department of Physics and Astronomy, The University of New Mexico, Albuquerque, NM 87106, USA\\
$^{20}$Department of Aeronautics and Astronautics, Massachusetts Institute of Technology, 77 Massachusetts Avenue, Cambridge, MA 02139, USA \\
$^{21}$Department of Astrophysical Sciences, Princeton University, 4 Ivy Lane, Princeton, NJ 08544, USA \\
}

\bsp	
\label{lastpage}
\end{document}